\documentclass[pra,showpacs,twocolumn,superscriptaddress]{revtex4}

\usepackage{graphicx}
\usepackage{multirow}
\usepackage[usenames]{color}
\usepackage{nicefrac}
\usepackage{amsmath}
\usepackage{bm}
\usepackage{booktabs}

\begin{document}

\title{Double-target BEC atomtronic rotation sensor}

\author{Oluwatobi Adeniji}
\affiliation{Department of Biochemistry, Chemistry, and Physics, Georgia Southern University, Statesboro, GA 30460-8031 USA}

\author{Charles Henry} 
\affiliation{Department of Biochemistry, Chemistry, and Physics, Georgia Southern University, Statesboro, GA 30460-8031 USA}

\author{Stephen Thomas} 
\affiliation{Department of Biochemistry, Chemistry, and Physics, Georgia Southern University, Statesboro, GA 30460-8031 USA}

\author{Robert Colson Sapp} 
\affiliation{Department of Biochemistry, Chemistry, and Physics, Georgia Southern University, Statesboro, GA 30460-8031 USA}

\author{Anish Goyal} 
\affiliation{Department of Biochemistry, Chemistry, and Physics, Georgia Southern University, Statesboro, GA 30460-8031 USA}

\author{Charles W.\ Clark}
\affiliation{Joint Quantum Institute, National Institute of Standards 
and Technology and the University of Maryland, Gaithersburg, MD 20899, USA}

\author{Mark Edwards}
\affiliation{Department of Biochemistry, Chemistry, and Physics, Georgia Southern University, Statesboro, GA 30460-8031 USA}
\affiliation{Joint Quantum Institute, National Institute of Standards 
and Technology and the University of Maryland, Gaithersburg, MD 20899, USA}

\date{\today}

\begin{abstract}
We present a proof-of-concept design for an atomtronic rotation sensor consisting of an array of ``double-target'' Bose-Einstein condensates (BECs). A ``target'' BEC is a disk-shaped condensate surrounded by a concentric ring-shaped condensate.  A ``double-target'' BEC is two adjacent target BECs whose ring condensates partially overlap. The sensor consists of an $n\times m$ array of these double-target BECs.  The measurement of the frame rotation speed, $\Omega_{R}$, is carried out by creating the array of double-target BECs (setup step), inducing one unit of quantized flow in the top ring of each member of the array (initialization step), applying potential barriers in the overlap region of each member (measurement step), and observing whether the induced flow is transferred from the top to the bottom ring in each member (readout step). We describe a set of simulations showing that a single instance of a double-target BEC behaves in a way that enables the efficient operation of an $n\times m$ array for measuring $\Omega_{R}$. As an example of sensor operation we present a simulation showing that a 2$\times$2 array can be designed to measure $\Omega_{R}$ in a \textcolor{black}{user-specified} range. 
\end{abstract}


\maketitle

\section{Atomtronic Sensor for Precision Navigation}
\label{atomtronic_sensor}

Inertial navigation systems (INS), which include gyroscopes and accelerometers, are used in many applications such as commercial aircraft and ocean-going vessels~\cite{navbook}. These systems enable moving vehicles to determine their position and orientation using the method of ``dead reckoning.''  In this method, the vehicle's location and orientation are determined relative to a known reference point by integrating the vehicle's acceleration and angular velocity since it left the reference point. This method enables one to determine position and orientation continuously without needing to receive external signals from satellite navigation systems such as the Global Positioning System (GPS)\,\cite{navbook, ins_conference}.

The ability to navigate without outside help is valuable because systems like the GPS can be jammed or spoofed by bad actors. Furthermore, GPS is not always available as, for example, deep under the ocean. Thus, the development of on-board inertial sensors that can operate for \textcolor{black}{times long compared to trip times} in the absence of external signals, \textcolor{black}{such as GPS,} constitutes a important research problem. It would be a significant achievement if sensors were capable of operating accurately for long periods of time~\cite{El-Sheimy}.

All inertial navigation systems have parameters that must be calibrated. These \textcolor{black}{parameters} include biases, scale factors, and misalignment of their various components~\cite{navbook}.  The values of these parameters are not always constant and can slowly change over the operating period of the INS device.  This is referred to as ``parameter drift.'' Parameter drift leads to errors in the measurement of accelerations and rotation velocities and this is a particular problem for navigation systems that are based on dead reckoning. Since position and orientation are determined by integrating rates of change over time, parameter drift causes the position and orientation errors to grow with time.  It is therefore useful for the INS to include a secondary system for measuring acceleration and rotation angular velocity more accurately, but at a lower sample rate to correct for errors incurred by parameter drift~\cite{navbook}.\

In this work, we propose a rotation sensor design based on an ``atomtronic'' system as a possible secondary measurement system.  An atomtronic system is loosely defined as an ultracold gas manipulated by laser light so that its behavior is analogous to an electronic system, except that it has a current of neutral atoms rather than electrons.  Such gases are generally cold and dense enough to be put in the Bose-Einstein condensate (BEC) state.

Recent advances in the optical manipulation of neutral atoms\,\cite{demarco_2008, hadzibabic_2012, boshier_2009, donatella_cassettari} have sparked experimental and theoretical interest in systems of Bose-Einstein-condensed atomic gases confined to a thin sheet in a horizontal plane. Cases where the BEC is confined within this plane to a closed-loop channel potential can be roughly analogous to electronic circuits. These systems are sometimes referred to as ``atom circuits'' and their study is part of the emergent field of atomtronics\,\cite{Amico_2017}.  

Interest in atom circuits derives in part from their potential for use in devices such as rotation sensors\,\cite{Amico_2017} suitable for precision navigation.  Proposed examples include devices that sense rotation via Sagnac interferometry\,\cite{PhysRevLett.124.120403, 2000Gustavson}, and those that act as analogs of Superconducting Quantum Interference Devices (SQUIDs), where rotation takes the place of magnetic flux\,\cite{boshier_2013a, Mathey2016, njp_paper, qs7}. Some implementations of these types of interferometers include a BEC confined in a ring geometry\,\cite{Bell2016, ring_circ_probe, sq3, sq7, sq8, 1st_ringBEC_current, 2nd_ringBEC_current, Kumar2016, hysteresis_nature_paper, resistive_flow_BEC}. In this work, we present a different idea for measuring frame rotation speed. 

This work was inspired by earlier work in double-ring geometries that showed that persistent currents could exist in a single ring without transferring to the other ring~\cite{Bland_2020} and that such currents are dependent on the rotation and acceleration of the rest frame of the double ring~\cite{PhysRevResearch.4.043171}. We also note that there an idea for measuring frame acceleration using the motion of vortices has been proposed~\cite{chaika2024}.

\begin{figure}
    \centering
    \includegraphics[scale=0.2]{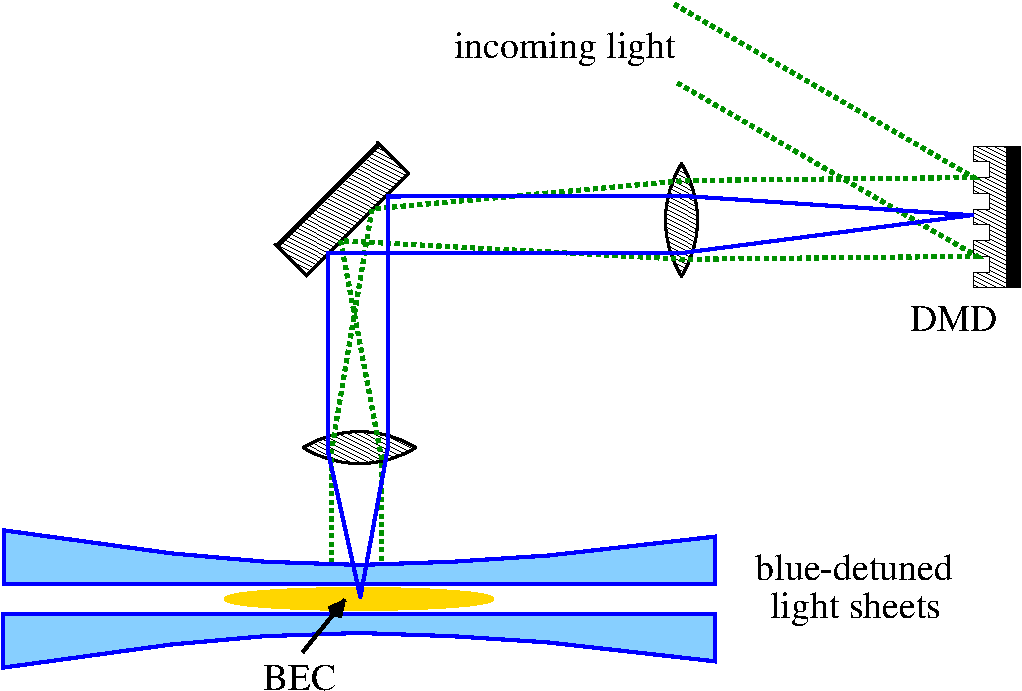}
    \caption{The general setup of the atomtronic system. A gas of neutral sodium atoms is confined to a thin sheet in the horizontal plane by a pair of blue-detuned light sheets~\cite{pethick_smith_2008}.  Within this plane an arbitrary 2D optical potential can be applied to these atoms using a digital micromirror device (DMD). This figure is drawn after Fig.\,2(a) of Ref.\,\cite{Kumar2016}}.
    \label{atomtronic_setup}
\end{figure}

\textcolor{black}{Here we consider a particular implementation of the general setup depicted in Fig.\,\ref{atomtronic_setup}. A gas of sodium atoms is squeezed into a horizontally oriented thin sheet by laser light. It is then subjected to a 2D potential within the plane that is arbitrary in space and time. A laser beam, whose photon energy is tuned to the blue of an atomic transition, acts as a repulsive potential on the center-of-mass motion of the atom.  The potential is proportional to the laser intensity, which can vary in space and time. When the laser is tuned to the red of the transition, the laser intensity acts as an attractive potential~\cite{pethick_smith_2008}.  Thus, as can be seen in the figure, two parallel, horizontal, blue-detuned light sheets squeeze the gas into the space between the sheets and support the gas against gravity.}

\textcolor{black}{A two-dimensional potential is projected onto the space between the light sheets by focusing an image created by a digital micromirror device (DMD).  A DMD is a rectangular array of tiny mirrors, each of which contributes a pixel of the image by either reflecting laser light or scattering it away. Laser light is shined on the mirror array and the reflected light is imaged into the space between the light sheets as shown in Fig.\,\ref{atomtronic_setup}.  Given the 20 kHz mirror array refresh rate, a nearly arbitrary 2D potential can be applied to the atoms in the gas trapped between the sheets.  Such setups now exist in many labs around the world~\cite{dmd_reference,roadmap}.}

\textcolor{black}{We assume here that the gas of atoms has been confined as described above. We furthermore assume that this gas has been cooled such that a Bose-Einstein condensate \textcolor{black}{has} formed and that all of the gas atoms are in the condensate.  The potentials we refer to here are the 2D potentials created by the image projected by the DMD device.  The full potential experienced by the atoms, and that governs the center-of-mass motion of the atoms, is this 2D potential plus a harmonic oscillator potential in the vertical ($z$-axis) direction.}

\section{Proposed Atomtronic Rotation Sensor}
\label{proposed_sensor}

We propose an atomtronic rotation sensor design that consists of an $n\times m$ array of Bose-Einstein condensates, each of which is confined in a quasi-2D double-target potential. Here we present an overview of this rotation sensor and its operation.

\begin{figure}
\centering
\includegraphics[scale=0.25]{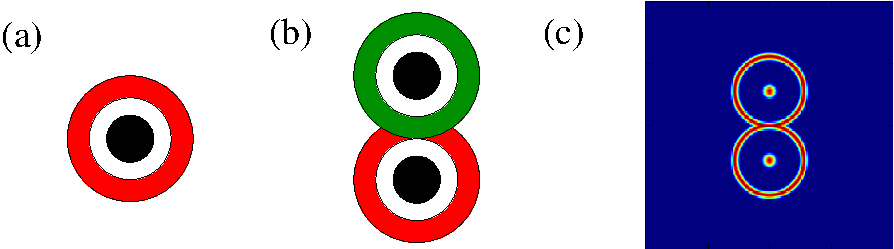}
\caption{(a) A single target potential supports a condensate having a central disk with a surrounding ring: a target BEC. (b)  Here there are two target BECs whose rings overlap: a double-target BEC.  The green ring has one unit of flow and the red ring has no flow. (c) Simulation view of a double-target BEC density. The colors represent density contours.}
\label{single_target_pair}
\end{figure}

Each 2D ``target'' potential consists of a central well surrounded by a concentric ring-shaped channel potential as shown in Fig.\,\ref{single_target_pair}. A ring and disk BEC, which we will call a ``target BEC,'' can be formed in this potential. This target configuration was realized in the matter-wave interferometry experiment reported in  Ref.\,\cite{PhysRevX.4.031052}.

A cartoon top view of a target BEC is shown in Fig.\,\ref{single_target_pair}(a).  The disk BEC will be used as a stationary phase reference to determine the flow in the outer ring by interferometry. It is possible to induce one unit of quantized flow (where the average angular momentum per particle is $\hbar$) in the ring BEC by either stirring or phase imprint~\cite{PhysRevX.4.031052, roadmap}. The ring condensate shown there is colored red indicating that the condensate is not flowing.

A ``double-target'' potential consists of a figure-eight channel potential of uniform depth surrounding the two disks. This resembles a superposition of
two adjacent target potentials that have a region of overlap as shown in Fig.\,\ref{single_target_pair}(b) and motivates our name for the potential. 

We assume that a ``double-target BEC'' can be formed in this potential, as it is consistent with the current state of experimental art\,\cite{roadmap}. In Fig.\,\ref{single_target_pair}(b) the top ring is colored green to indicate that there is one unit of quantized flow in its BEC. The bottom ring is colored red to indicate that its BEC has zero units of quantized flow. 

Our simulations show that it is possible to induce one unit of flow in the upper ring and no units of flow in the lower ring. In this case the net change in phase around a closed loop contained in the upper ring is $2\pi$ while the net change in phase around a loop in the lower ring vanishes. This condition will persist as long as the potential is unchanging. \textcolor{black}{Figure \ref{single_target_pair}(c) shows an example of the double-target BEC phase distribution from our Gross-Pitaevskii simulations.}

\begin{figure}
\centering
\includegraphics[scale=0.25]{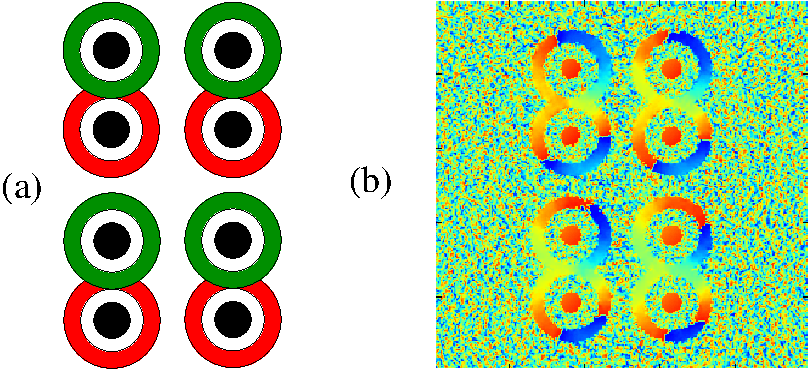}
\caption{(a) A cartoon picture of a 2$\times$2 array of double-target BECs. (b) Simulation view of the BEC phase in the same potential. The color of each pixel represents the phase of the wave function at that point.}
\label{dtap_example}
\end{figure}

An $n\times m$ ``double-target-array'' potential (DTAP) consists of an $n$-row by $m$-column array of double-target potentials.  When the gas is cooled in this potential an array of double-target BECs is formed. A cartoon picture of BECs formed in a 2$\times$2 DTAP is shown in Fig.\,\ref{dtap_example}(a).  The phase view from one of our simulations is shown in Fig.\,\ref{dtap_example}(b).  

We note that the solution of the rotating-frame Gross-Pitaevskii equation (RFGPE) shown in Fig.\,\ref{dtap_example}(b) is obtained by numerical solution at each point of a grid that covers the entire region displayed. The details of our RFGPE model are given in Appendix\,\ref{sim_model}. We found that each member of the array behaves independently of the other members.

For this arrangement of double-target BECs to act as a rotation sensor we first assume that the rest frame of the DTAP is rotating at speed $\Omega_{R}$ with respect to the ``fixed stars.''  The purpose of our rotation sensor is to measure $\Omega_{R}$. To see how this can happen we first describe the behavior of a single instance of a double-target BEC.

Our simulations show that, if we create a double-target BEC and induce one unit of flow in the top ring, as shown in Fig.\,\ref{single_target_pair}(b), in a non-rotating frame ($\Omega_{R}=0$), then the flow will remain in the top ring indefinitely. This behavior is expected from the vortex model of circulation. If the top ring has flow, then there is a vortex trapped inside that ring and transfer can't occur unless the vortex has a path to travel to the other ring. This path can be created by applying a barrier potential in the overlap region between the two rings.  We find that transfer can occur via the combined effects of frame rotation and the presence of a barrier erected across the overlap region of the two rings.

If a Gaussian-shaped potential barrier is turned on and then off across the overlap region with a sufficiently strong maximum height (we call it $U_{b,max}$), this can cause the flow to transfer from the top ring to the bottom ring.  We find that, when $\Omega_{R}=0$,  there is a critical barrier strength, $U_{b,max,c}$, such that the flow does not transfer when $U_{b,max} < U_{b,max,c}$.  

\textcolor{black}{Thus, if we turn on a barrier where $U_{b,max}$ is below $U_{b,max,c}$, then the combined effect of barrier and frame rotation can cause the flow to transfer between rings. Thus, whether or not the flow transfers can now be made sensitive to the presence of a small non-zero rotation speed $\Omega_{R}$.} It may, therefore, be possible to design rotation sensors of this type that are sensitive to arbitrarily small frame rotation speeds, subject to noise limitations.

Next, we keep the value of $U_{b,max}$ fixed, but less than $U_{b,max,c}$, and look at what happens when we turn on the barrier for non-zero values of $\Omega_{R}$.  We find that there is a critical value, $\Omega_{c}$, of the rotation speed where the flow doesn't transfer when $\Omega_{R} < \Omega_{c}$ and does transfer when $\Omega_{R} > \Omega_{c}$.  

Thus, when we turn the barrier on and off for an unknown value of $\Omega_{R}$, if the flow transfers, then it must be that $\Omega_{c} < \Omega_{R}$, and the critical value, $\Omega_{c}$, puts a lower bound on $\Omega_{R}$.  If the flow does not transfer, then then it must be that $\Omega_{c} > \Omega_{R}$ and the critical value puts an upper bound on $\Omega_{R}$. 

\textcolor{black}
{It is worth noting here that, we are assuming that the direction of rotation of our rotating frame, relative to the $\Omega_{R}=0$ frame, is known. An example of this might be the measurement of the Earth's rotation speed.  In our case, the imprinted flow direction is opposite that of the frame rotation.} This is not an essential feature and we will say more about it later.

Now suppose that we have two double-target BECs on which barriers of different heights, $U_{b,max,1}$ (corresponding to critical value $\Omega_{c1}$) and $U_{b,max,2}$ (corresponding to critical value $\Omega_{c2} > \Omega_{c1}$) were turned on and off.  If, for example, we found that the circulation for the $U_{b,max,1}$ BEC transferred, then we would have $\Omega_{c1} < \Omega_{R}$. If also the circulation for the $U_{b,max,2}$ did not transfer, then we would also have $\Omega_{R} < \Omega_{c2}$.  In this case, where the circulation of one double-target BEC did transfer and the other did not, we could conclude that $\Omega_{c1} < \Omega_{R} < \Omega_{c2}$ thus placing both a lower and an upper bound on $\Omega_{R}$.  This would then constitute a measurement of $\Omega_{R}$.

This sensor idea will only work if there is a one-to-one correspondence between the maximum barrier height, $U_{b,max}$, and the critical rotation speed, $\Omega_{c}$ where transfer doesn't occur if $\Omega_{R} < \Omega_{c}$ and does occur if $\Omega_{R} > \Omega_{c}$.  If this is the case, then it should be possible to put narrow bounds on the frame rotation speed $\Omega_{R}$ by using an array of double-target BECs and raising barriers in the overlap region of each one where the $U_{b,max}$ for each double-target BEC is different.

In what follows we will also assume that the array of double-target BECs lies far away from the rotation axis.  We will show that the transfer behavior (that is whether or not the circulation transfers) of a 1$\times$1 array does depend on how far away it is from the axis.  However, we will also show that, once the distance to the axis becomes much larger than the array size, the transfer behavior stays the same no matter how much further away it gets.  

We will also show that, far from the axis, the transfer behavior of each individual member of an $n\times m$ double-target BEC array is the same as if it were by itself and is not influenced by the presence of the other members of the array. This important property will enable us to study the transfer behavior of a 1$\times$1 array, far from the rotation axis, and then use this data to design an $n\times m$ array (by setting a unique  value of $U_{b,max}$ for each array member) that will be sensitive to $\Omega_{R}$ in a specified range.  We will demonstrate a proof-of-concept example of this process below.

Finally, we will show how the transfer behavior of each member of the double-target BEC array can be measured from a single density image. By turning off the double-target potentials and allowing the ring BECs to overlap the phase-reference disk BECs, the resulting interference patterns can probe whether circulation was present in the ring or not.  Using laser light to measure an absorption image is the standard method for probing ultracold-atom systems~\cite{dmd_reference}.

This paper describes a study of the operation and design of an atomtronic rotation sensor consisting of an $n\times m$ array of double-target BECs. In Section\,\ref{simulations}, we will describe the conditions common to all of the simulations. In Section\,\ref{cycle_chapter}, we present the full operating cycle of the proposed DTAP sensor including setup, initialization, measurement, and readout. Section\,\ref{sim_results} contains the results of our simulations of sensor behavior. We demonstrate the transfer behavior of one double-target BEC as described above. \textcolor{black}{Also,} we use this result to design a 2$\times$2 array which is sensitive to values of $\Omega_{R}$ within a \textcolor{black}{user-specified} range. We then present the results of simulating a 2$\times$2 sensor array using this design. Finally, we show how the transfer behavior of the array can be determined from a single density image and compare this with the results of computing the winding number. Section\,\ref{summary}, presents a summary and future work.

\section{Simulation Model and Conditions}
\label{simulations}

The results presented later in this work are based on a set of simulations that model the setup, initialization, measurement, and readout steps of the operation of the rotation sensor.  These steps will be described in more detail in the next section. The rotating-frame Gross-Pitaevskii equation model used for simulating sensor behavior is described in detail in App.\,\ref{sim_model} and the details of all the potentials introduced below are contained in App.\,\ref{potentials}.  Here we describe the characteristics that are common to all of the simulations. 

In all of the simulations, the atomic species was $^{23}$Na and there were $N_{\rm atoms}=166,667$ atoms per double-target pair.  The ring mid-track radius was taken to be $R_{r}=22.5\,\mu$m and the ring width was $w_{r}=5.0\,\mu$m.  The disk width was $w_{d}=10.0\,\mu$m. The depth of the double-target potentials was $V_{0}=209.2$ nK. The widths of the Gaussian barriers were $w_{\parallel} = w_{\perp} = 11.25\,\mu$m.  \textcolor{black}{These numbers were chosen because they are similar to characteristics of previous experiments on ring BECs~\cite{exp_universe, Kumar2016, hysteresis_nature_paper, resistive_flow_BEC}.}

\begin{figure}
\centering
\includegraphics[scale=0.30]{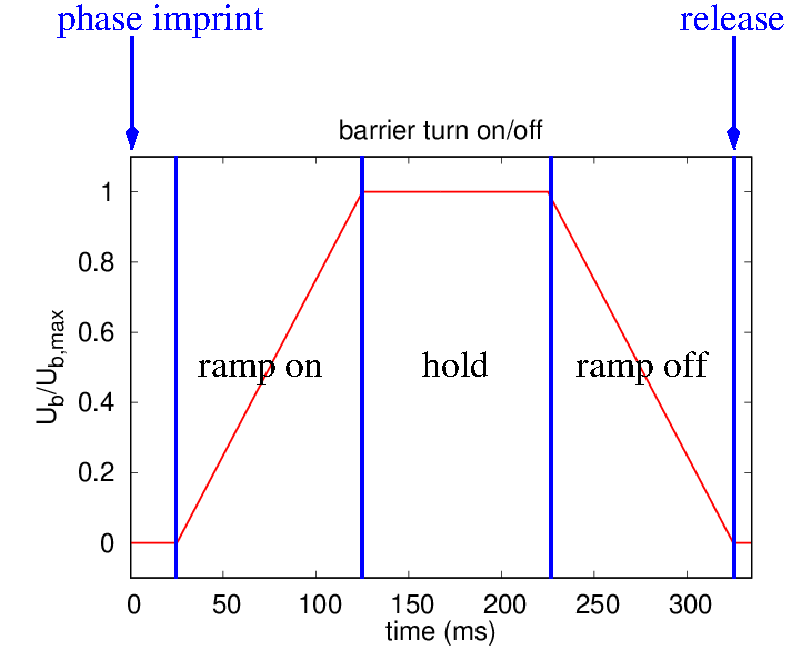}
\caption{\textcolor{black}{Normalized energy height of the barriers versus time. This time profile was the same for all simulations.} For the first 25 ms, the barriers are off, to allow the phase-imprint to settle. The barriers are then turned on linearly over a time of 100 ms, held constant for a further 100 ms, ramped linearly off over 100 ms, and then all potentials are turned off, and system evolution proceeds for another 5 ms.}
\label{barrier_on_off}
\end{figure}

The first step of the simulation was to compute the initial condensate wave function.  This is equivalent to the evaporative cooling step in an experiment in which the BECs are initially formed.  In our simulations, this was done by integrating the time-dependent RFGPE in imaginary time.  The result of this is a solution of the time-independent RFGPE.  The condensate wave function that results produces a density like that shown in Fig.\,\ref{single_target_pair}(c).

\begin{figure*}
\centering
\includegraphics[scale=0.28]{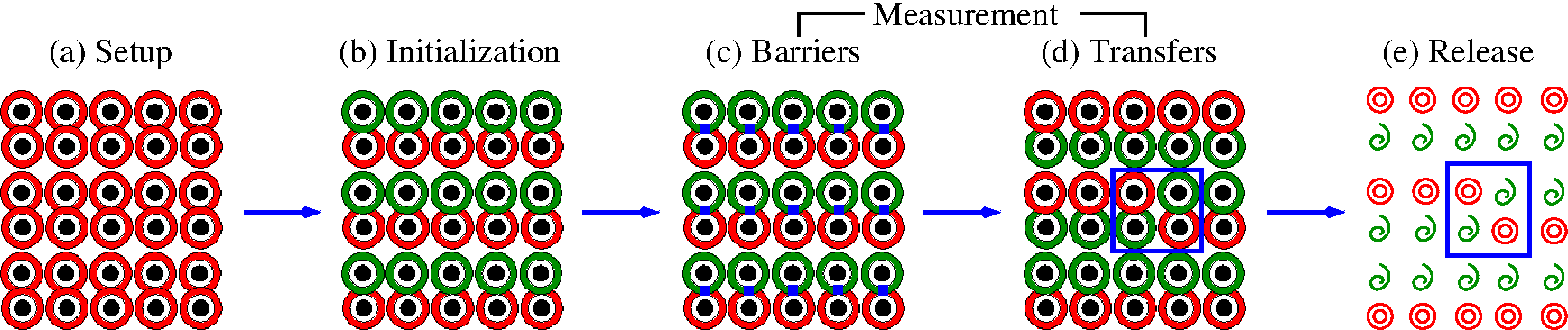}
\caption{Cartoon of the proposed rotation sensor operating cycle. (a) Setup: The rectangular array of double-target BECs is created with no circulation (red rings) in any of the ring BECs. (b) Initialization: Flow is induced on the top ring of each double-target BEC in the array. (c),(d) Measurement: Barriers of different energy heights, $U_{b,max}$, are applied to the overlap region of each double-target BEC with the highest $U_{b,max}$ applied to the upper-left double-target BEC to the lowest value in the lower right. The barriers are turned on and off as described in Fig.\,\ref{barrier_on_off}. The result is that the circulation has transferred from top to to bottom in some of the double-target BECs but not in others. The crossover from transfer to non-transfer is shown in the blue rectangle. (e) Release: The traps are all turned off. Each ring condensate fills in to overlap with its disk condensate phase reference. Rings without flow show an interference pattern with concentric rings and rings with flow show a spiral interference pattern. A single image can now be taken to determine lower and upper bounds on $\Omega_{R}$.}
\label{sensor_operation}
\end{figure*}

The next step in the simulation was to induce flow in the top ring of each double-target BEC in the array.  Experimentally this can be done using laser light~\cite{dmd_reference}.  In our simulation, we do this by multiplying the condensate wave function by the factor $e^{i\phi}$ at the site of each top ring, where $\phi$ is the angle measured from the $+x$ axis of an $xy$ coordinate system centered on the top ring.  This imposes a phase gradient on the wave function and induces a flow.  Phase imprinting is commonplace now because of the presence of DMD systems in many BEC labs around the world\,\cite{GAUTHIER20211}. The system was then allowed to evolve for a 25 ms wait period while the sudden application of phase gradient settles.  This is also done in experiments.

The next step was to turn the barrier potentials on and off. The time dependence of the barriers was the same for all simulations and across all of the double-target BECs in the array.  This time dependence is shown in Fig.\,\ref{barrier_on_off}. All of the barriers were first linearly ramped up to their maximum energy height, $U_{b,max}$, over 100 ms, held constant for 100 ms, and then linearly ramped off over a final 100 ms. 

The final step in our simulations was to turn off the ring and disk potentials holding the condensates in place and allowing the system to evolve for another 5 ms. During this time the rings start to fill in while the disks start to expand outward. These quickly overlap producing an interference pattern. If there was no flow in the ring, the interference pattern is a set of concentric rings. If flow was present, the pattern is a spiral.  In this way, the rings in which flow was present at the moment of release can be detected with a single density image.

\section{Rotation sensor operating cycle}
\label{cycle_chapter}

We now outline the operating cycle in which the double-target BEC array rotation sensor determines the speed of the rotating frame.  These steps are illustrated in Fig.\,\ref{sensor_operation} as: (a) setup, (b) initialization, (c) and (d) measurement, and (e) readout.

(a) Setup: a rectangular array of double-target BECs is created, but there is no flow present in any of the rings.  A cartoon illustration for a $3\times 5$ array is shown in Fig.\,\ref{sensor_operation}(a).  The red coloring of the rings means that there is no flow in the rings. The black disks designate the zero-flow, phase-reference BECs. These wil be used in the readout step.

(b) Initialization: one unit of quantized flow is induced in the top ring of each double-target BEC by phase imprint. The bottom ring of each double-target BEC has no flow. At this point the sensor is initialized so that it is ready to make a measurement. This condition is shown in Fig.\,\ref{sensor_operation}(b).


(c) Measurement (barriers): a barrier potential is raised across the overlap region of each double-target BEC. The maximum value of the barrier energy height, $U_{b,max}$, is different for each double-target BEC across the array. The time schedule for turning the barriers on and off is the same for all double-target BECs as shown in Fig.\,\ref{barrier_on_off}.  

During this step, the target-BEC-pair array appears as shown in Fig.\, \ref{sensor_operation}(c) where the barriers are represented as blue rectangles. In the particular example discussed here, we used Gaussian barriers.


(d) Measurement (transfers): After barriers have been turned on and off, flow has transferred from the top ring to the bottom ring in some of the double-target BECs, while in others there was no transfer.  This situation is illustrated in Fig.\,\ref{sensor_operation}(d).  

(e) Readout:  The target potential confining all of the condensates is turned off for a brief period after which an image of the final state of the system is taken.  A cartoon of one possible result of this process is shown in Fig.\,\ref{sensor_operation}(e).  

After the turnoff, the rings fill in while the disks expand and their overlap displays an interference pattern.  If flow is present in a given ring, then the interference pattern is a single spiral.  If there is no flow in the ring, then the interference is a set of concentric rings. 

It is important to note that it is not necessary to use double-``target'' BECs, i.e., ring BECs surrounding disk BECs for readout of the flow in a ring. A simpler possibility would be to just have double-rings.  Readout would proceed as described. Rings with flow would fill in but leave a hole in the center while rings with no flow would fill in completely\,\cite{ring_circ_probe}.  A readout step like this would be slower than the target-BEC idea.

As our simulations will show, each double-target BEC has a different critical transfer rotation speed, $\Omega_{c}$.  \textcolor{black}{Thus, at the end of the cycle, there will be a set of double-target BECs for which flow did not transfer, and the largest $\Omega_{c}$ among this group puts the largest lower bound on $\Omega_{R}$.  All the other double-target BECs did transfer and the lowest value of $\Omega_{c}$ among this group puts the smallest upper bound on $\Omega_{R}$.} The blue rectangle shown in Fig.\,\ref{sensor_operation}(e) encloses the double-target BECs which put the closest lower and upper bounds on $\Omega_{R}$.

It is clear that a necessary characteristic of a double-target BEC for the sensor operation to work is that, once the barrier is applied to the double-target BEC, the flow induced in the top ring BEC will transfer to the bottom ring if the rotating-frame speed, $\Omega_{R}$,  exceeds a critical threshold value, $\Omega_{c}$, and will not transfer if the frame speed is less than this value.  Our simulation studies, described next, will show that double-target BECs actually can exhibit this behavior. 


\section{Simulation results}
\label{sim_results}

In this section, we present the results of our simulations. The goal of these simulations was to show that it is possible to design an array of double-target BECs that can used to measure $\Omega_{R}$ within a given range.


In Section\,\ref{zero_rot_studies} we present the results of studying the transfer behavior of a $1\times1$ array when the rotation speed is zero ($\Omega_{R}=0$), and for successively increasing values of the maximum barrier strength, $U_{b,max}$. We show here that no transfer occurs for $U_{b,max}$ less than a critical value $U_{b,max,c}$.  

Next we performed a series of simulations (reported in Section\,\ref{fixed_Ubmax_studies}) where  $U_{b,max}<U_{b,max,c}$ is fixed, but the value of $\Omega_{R}$ is increased.  Here we show that, for a fixed value of $U_{b,max}$, there is a critical value of $\Omega_{R}$ (which we will call $\Omega_{c}$), such that transfer does not occur when $\Omega_{R}<\Omega_{c}$ and transfer does occur when $\Omega_{R}>\Omega_{c}$.  We also show that each value of $U_{b,max}$ corresponds to a unique value of $\Omega_{c}$.  Finally, we will see that the values of $\Omega_{c}$ get smaller as $U_{b,max}$ approaches $U_{b,max,c}$.

We then use the results of the simulations that establish the correspondence of $U_{b,max}$ with $\Omega_{c}$ to design a rotation sensor consisting of a $2\times2$ array of double-target BECs that can measure $\Omega_{R}$ within a specified range. The ``design'' activity simply involves designating the value of $U_{b,max}$ for each member of the $2\times2$ array.  

Finally, in Section\,\ref{sensor_design}, we present the results of a simulation of our designed rotation for a chosen value of $\Omega_{R}$.  We show that the transfer behavior of the simulated $2\times2$ array agrees with our predicted behavior.  We also show how the transfer behavior of all members of the $2\times2$ array of double-targets can be inferred from a single density image.  \textcolor{black}{We also present a comparison of the density-image results with our computed values of the winding number.}

\subsection{Zero-rotation study}
\label{zero_rot_studies}

We now turn to a study of the transfer behavior of a $1\times1$ double-target BEC array whose center is far from the rotation axis in a non-rotating reference frame.  In these simulations we set $\Omega_{R}=0$ and perform a series of runs with progressively larger and larger values of $U_{b,max}$.  After each run, we look at the transfer behavior of the double-target BEC.

\begin{figure}
    \centering
    \includegraphics[scale=0.34]{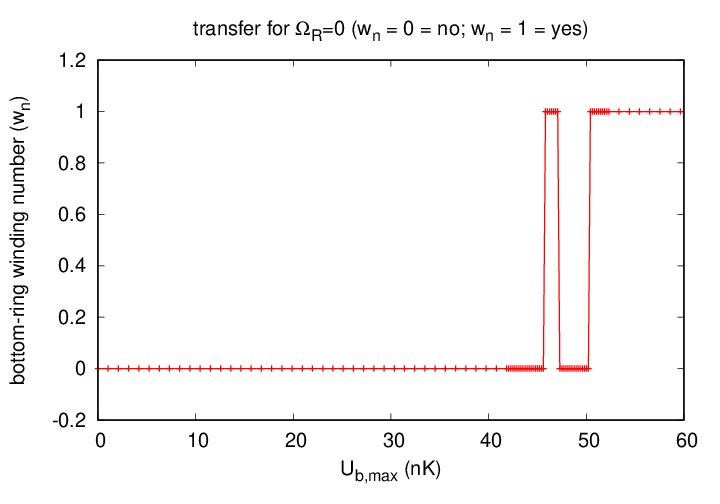}
    \caption{The winding number, $w_{n}$, of the bottom ring of the double-target BEC after the barrier has been turned on and off.  The frame is non-rotating, $\Omega_{R}=0$ rad/sec. If $w_{n}=0$, then transfer did not occur and, if $w_{n}=1$, then transfer did occur.}
    \label{zero_rot_graph}
\end{figure}

The result of this study is shown in Fig.\,\ref{zero_rot_graph}. This figure shows a plot of the winding number, $w_{n}$, of the bottom ring of the double-target BEC after the barrier has been turned on and off.  If no transfer occurred, then $w_{n}=0$ for the bottom ring, and if transfer did occur, then $w_{n}=1$.

This graph shows that, at zero frame-rotation speed, there is no transfer when $U_{b,max}$ is less than a critical value, $U_{b,max,c}=45.8$ nK. If $U_{b,max} > U_{b,max,c}$ then it is possible for the barrier to effect transfer by itself.  Thus, we will only consider conditions where $U_{b,max} < U_{b,max,c}$ where transfer can only occur through a combination of the barrier and non-zero frame-rotation speed.

One possible implication of this result is that, if we set $U_{b,max}$ to a value just below $U_{b,max,c}$ then, even though no transfer happens when $\Omega_{R}=0$, transfer might occur for a small value of $\Omega_{R}$.  Thus, turning a barrier on and off with $U_{b,max}$ just below $U_{b,max,c}$ can make the system sensitive to the rotating-frame speed.

\begin{figure}[ht]
\centering
\includegraphics[scale=0.3]{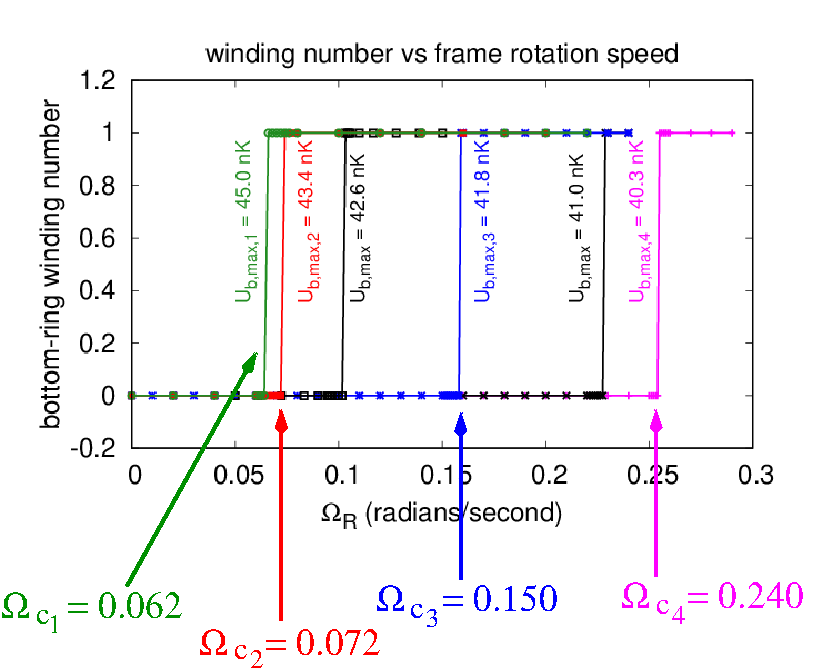}
\caption{Bottom-ring winding number, $w_{n}$ as a function of rotation speed, $\Omega_{R}$ for six values of $U_{b,max}$. Four of these values are chosen for the design of the 2x2 rotation sensor. The associated critical rotation speeds needed for transfer for these values of $U_{b,max}$ are also shown.}
    \label{1x1_wn_vs_omegaR}
\end{figure}

\subsection{Studies with fixed $U_{b,max}$ and varying $\Omega_{R}$}
\label{fixed_Ubmax_studies}

In the next study, we kept $U_{b,max}$ fixed and less than $U_{b,max,c}$. We then conducted simulations for different values of $\Omega_{R}$. We started with $\Omega_{R}=0$ and performed simulations with progressively larger and larger values of $\Omega_{R}$.  In each simulation we looked at the transfer behavior of the double-target BEC.  In all of these simulations the double-target BEC center was located far away from the rotation axis.

The results of these runs are depicted in Fig.\,\ref{1x1_wn_vs_omegaR}.  This figure shows plots of the final winding number of the bottom ring of the double-target BEC as a function of the frame rotation speed, $\Omega_{R}$.  Each curve in the plot corresponds to a series of simulations where the value of $U_{b,max}$ is kept fixed and $\Omega_{R}$ is varied.  There are six curves on this plot, each for a different value of $U_{b,max}$.  Since we are showing the final winding number of the bottom ring, a winding number of zero means that transfer did not occur and a winding number of one means transfer did occur.

There are several interesting features of this plot.  The first notable feature is that, for fixed $U_{b,max}$ there is no transfer for small values of $\Omega_{R}$. However, above a critical value where $\Omega_{R}=\Omega_{c}$, the flow will transfer.  Next, we note that the critical rotation speed above which transfer happens depends on the value of $U_{b,max}$.  The correspondence between $\Omega_{c}$ and $U_{b,max}$ is shown on the plot and also in Table\,\ref{table1}.

We see here that, for a given value of $U_{b,max}$, there is a unique critical value of the frame-rotation speed, $\Omega_{R}=\Omega_{c}$, where transfer begins to occur for each value of $U_{b,max}$.  Furthermore, we can see that as $U_{b,max}$ increases, $\Omega_{c}$ decreases.  

This can be understood qualitatively. For $U_{b,max} < U_{b,max,c}$ the barrier is unable, by itself, to cause transfer. However, transfer can take place through the combined action of the barrier and the rotating-frame speed. The closer $U_{b,max}$ is to $U_{b,max,c}$ the addition of a smaller value of $\Omega_{R}$ will be able to effect transfer.

We conclude from this data that there is a one-to-one correspondence between the value of $U_{b,max}$ and the value of $\Omega_{c}$.  We can use this result to design an $n\times m$ double-target array where each successive double-target has a successively smaller value of $U_{b,max}$ and thus a successively larger value of $\Omega_{c}$. 

As a proof-of-concept, we will design a $2\times 2$ array in the next section for measuring $\Omega_{R}$ within a specified range and then simulate the actual behavior of the designed rotation sensor to compare with the expected result.

\begin{figure*}
    \centering
    \includegraphics[scale=0.49]{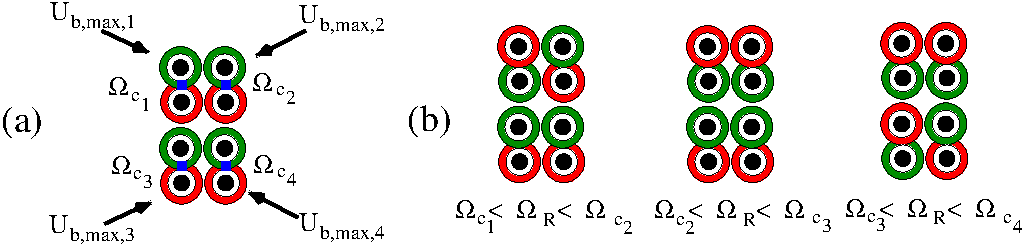}
    \caption{Design of a 2x2 double-target BEC rotation sensor. (a) Each double-target BEC has a unique maximum barrier energy height, and $U_{b,max}$ and associated critical frame-rotation speed for transfer, $\Omega_{c}$. (b) The possible transfer states after application of the barriers and the lower and upper bounds placed on the frame rotation speed, $\Omega_{R}$. }
    \label{2x2_design}
\end{figure*}

\subsection{Design of a $2\times2$ double-target BEC array rotation sensor}
\label{sensor_design}

In this section, we describe the procedure of designing a $2\times2$ double-target BEC array rotation sensor to measure the unknown value of $\Omega_{R}$. An illustration of this array is shown in Fig.\,\ref{2x2_design} where the double targets are numbered 1 (upper left), 2 (upper right), 3 (lower left) and 4 (lower right). This figure also shows the values of $U_{b,max,i}$ and $\Omega_{c_{i}}$ where $i=1,\dots,4$ associated with each double target.  

We want to choose the values of the maximum barrier heights so that $\Omega_{c_{1}}<\Omega_{c_{2}}<\Omega_{c_{3}}<\Omega_{c_{4}}$ so that the resulting sensor can measure the rotating-frame speed in the range $\Omega_{c_{1}} < \Omega_{R} < \Omega_{c_{4}}$. To ``design'' the double-target BEC array rotation sensor means that we specify the number, $N$, of double-target BECs in the array, and we specify the values of $U_{b,max,i}\ i=1,\dots,N$ that will be applied to the array during the measurement step such that 

Our design is illustrated in Fig.\,\ref{2x2_design}(a).  We show there a $2\times2$ double-target BEC array and set the values of $U_{b,max}$ as labeled there.  We use the plot in Fig.\,\ref{1x1_wn_vs_omegaR} to find suitable values for $U_{b,max}$. As shown there we set $U_{b,max}=U_{max,1}=45.0$ nK, $U_{max,2}=43.4$ nK, $U_{max,3}=41.8$ nK, and $U_{max,4}=40.3$ nK.  With this choice each double-target BEC will thus have a unique value of $\Omega_{c}$. These are given in Table\,\ref{table1}.

\begin{table}[ht]
    \centering
    \begin{tabular}{c|c}
    \toprule
     $U_{b,max}$ (nK)    &  $\Omega_{c}$ (rad/sec)\\
\midrule
    45.0 & 0.062 \\
    43.4 & 0.072 \\
    41.8 & 0.150 \\
    40.3 & 0.240
    \end{tabular}
    \caption{Maximum barrier height, $U_{b,max}$, versus critical frame rotation speed, $\Omega_{c}$, where flow transfer begins.}
    \label{table1}
\end{table}

For this design we can now state what measurement outcomes we can expect and what they mean. This is illustrated in Fig.\,\ref{2x2_design}(b).  First, it's clear that the flow of each individual double-target BEC will either transfer or not transfer.  In this case, there are five possible transfer behaviors of the $2\times2$ double-target BEC array that we expect based on the results of the $1\times1$ double-target BEC array results presented earlier.  These are (1) none of the double-target BECs transfer in which case we expect that $0<\Omega_{R}<\Omega_{c_1}$, (2) only the upper-left double-target BEC transfers, in which case we expect that $\Omega_{c_1} < \Omega_{R} < \Omega_{c_2}$, (3) both the upper-left and upper-right double-target BECs transfer, in which case we expect that $\Omega_{c_2} < \Omega_{R} < \Omega_{c_3}$, (4) all of the double-target BECs transfer except the lower-right one, in which case we expect that $\Omega_{c_3} < \Omega_{R} < \Omega_{c_4}$, and finally (5) all of the double-target BECs transfer, in which case we expect that $\Omega_{R} > \Omega_{c_4}$.  

We note here that case (5) does not actually put both a lower and an upper bound on $\Omega_{R}$ and that case (1) only does so because we know that $U_{b,max} < U_{b,max,c}$ so that no transfer occurs for $\Omega_{R}=0$. Thus, we only expect our designed rotation sensor to put explicit lower and upper bounds on $\Omega_{R}$ for cases (2), (3), and (4). Thus our designed sensor can only provide definite measurements of the frame-rotation speed in the range $\Omega_{c_{1}}<\Omega_{R}<\Omega_{c_{4}}$.  These three cases are depicted in Fig.\,\ref{2x2_design} and are labeled with the value by which $\Omega_{R}$ is bounded.  

We remark here that our sensor does not display very high resolution but our goal here is just to show the proof-of-concept.  If the sensor does not work for low resolution then there is no reason to proceed further.  Higher resolution can be obtained by using a larger array.

\subsection{Simulation of the $2\times2$ double-target BEC array rotation sensor}
\label{sensor_simulation}

Finally, we present the results of a simulation of our designed rotation sensor.  We set up our $2\times2$ double-target BEC array with values of $U_{b,max}$ as described in the previous section, set the value of $\Omega_{R}=0.2$ rad/sec, and ran the simulation as described in Section\,\ref{simulations}.  For this value of $\Omega_{R}$ we have $\Omega_{c_3} < \Omega_{R} < \Omega_{c_4}$ which corresponds to case (4) described earlier. The result we expect is that all of the double-target BECs in the array will transfer except the lower-right one.

\begin{figure*}[htb]
    \centering
    \includegraphics[scale=0.15]{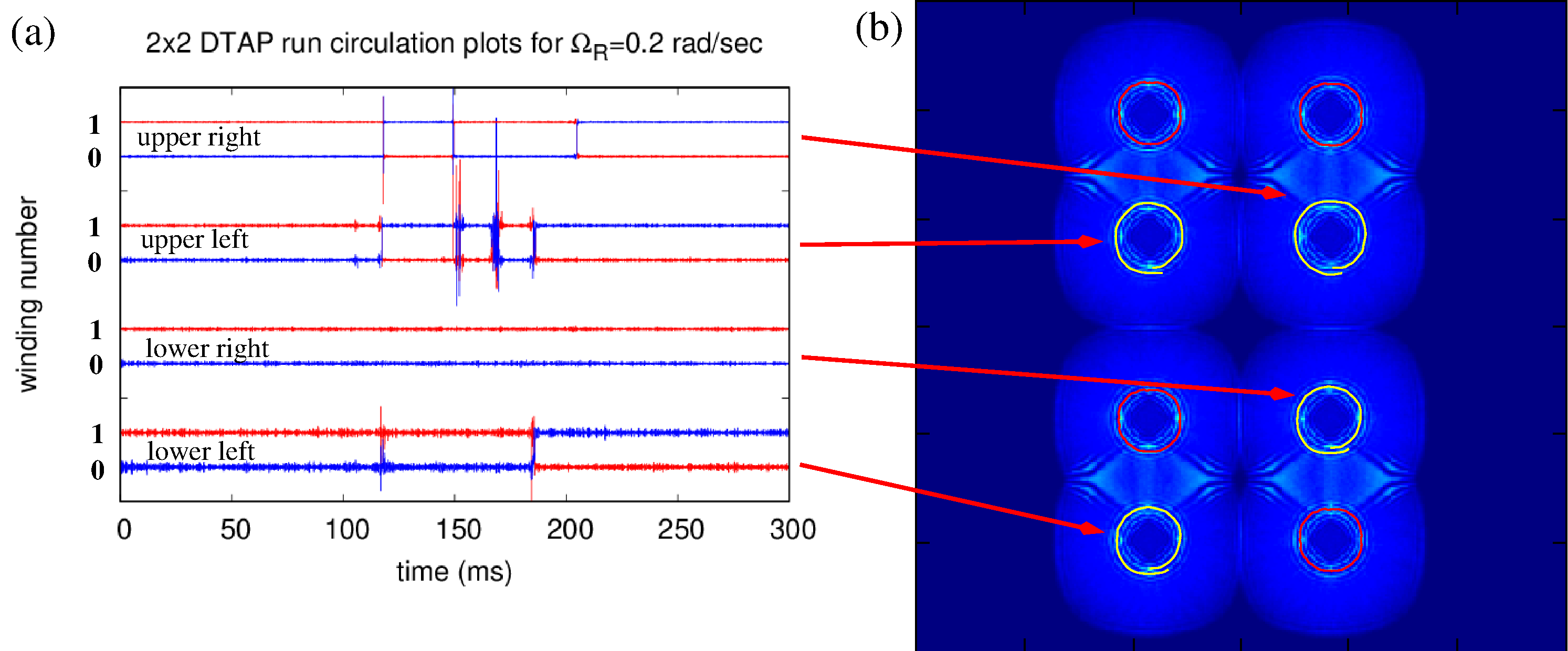}
    \caption{Results of simulating the designed $2\times2$ double-target array rotation sensor for $\Omega_R = 0.2$ rad/sec. (a) Winding numbers of the top (red) and bottom (blue) of the four double-target BECs as a function of time. (b) The integrated density of the array after a 4 ms expansion. The red circles highlight concentric ring interference patterns and the yellow spirals show the spiral interference patterns. The arrows point from the winding number plot for each double target to the spiral of the associated double target in the image. }
    \label{circulation_plot_2x2_simulation_for_0.2}
\end{figure*}

Figure\,\ref{circulation_plot_2x2_simulation_for_0.2} shows the results of the simulation. The left panel (\ref{circulation_plot_2x2_simulation_for_0.2}(a)) shows plots of the winding number of each double-target BEC as a function of time.  The right panel (\ref{circulation_plot_2x2_simulation_for_0.2}(b)) shows an image of the density after the potential is turned off and the system is allowed to expand for 4 ms.  Arrows have been drawn to point to the spiral pattern in each double target appearing in the density image.  

The density plot has been annotated with red circles to highlight the concentric ring interference patterns that indicate no flow. Ring BECs annotated with yellow spirals show density spirals indicating the presence of flow~\cite{spirals_paper}.  This shows that the transfer behavior can be determined experimentally by taking a single image after the measurement.

The figure shows that the outcome of our simulation of the designed $2\times2$ double target BEC array rotation sensor agrees with our expectation.  Both the final values of the winding numbers shown in the left panel and the circles and spirals seen in the density image show that flow transferred in all of the double-target BECs except the one on the lower right.  We take this example as an encouraging sign that our double-target BEC rotation sensor idea is feasible.

\section{Summary}
\label{summary}

In this paper, we presented a proposal for an atomtronic rotation sensor consisting of an array of double-target BECs.  The purpose of the sensor is to measure $\Omega_{R}$, the rotation speed of the sensor rest frame with respect to the fixed stars. The steps of operation of our proposed sensor include (1) setup, where an array of double-target BECs is formed; (2) initialization, where one unit of flow is imprinted on the top ring of each double-target BEC in the array; (3) measurement, where potential barriers of varying energy heights are turned on and then off across the overlap region of double-target BECs in the array; (4) readout, where the rings and disks are allowed to expand creating an interference pattern that determines the transfer behavior of each double-target BEC in the array.

Our simulations show (see App.\ref{rot_studies}) that the transfer behavior of both $1\times1$ and $2\times2$ arrays depends on the distance from the rotation axis to the array center and that this behavior stops changing when this distance becomes large compared to the size of the array.  Furthermore, we also found that the transfer behavior of each member of a $2\times2$ double-target BEC array is not affected by the presence of the other members of the array.  Thus we can study the transfer behavior of a $1\times1$ double-target array located far from the rotation axis and use this information to infer the behavior of an $n\times m$ double-target array.

We presented the results of a set of transfer behavior studies for a $1\times1$ double-target array located far from the rotation axis when $\Omega_{R}=0$ and when the maximum barrier height, $U_{b,max}$, was varied.  We found that, when $U_{b,max} < U_{b,max,c}$, no transfer occurred and, when $U_{b,max} > U_{b,max,c}$, transfer could occur.  Thus, by setting $U_{b,max} < U_{b,max,c}$ we could be sure that no transfer could be caused by the barrier by itself.

Next we studied the case where $U_{b,max} < U_{b,max,c}$ was fixed and $\Omega_{R}$ was varied.  We found that there was no transfer when $\Omega_{R}$ was less than a critical value, $\Omega_{c}$, but when $\Omega_{R}>\Omega_{c}$ transfer did happen. We concluded that, if we apply a barrier to the double-target BEC and find that transfer did not occur, then $\Omega_{R}<\Omega_{c}$ and this placed an upper bound on $\Omega_{R}$.  If transfer did occur, then $\Omega_{R}>\Omega_{c}$ and this placed a lower bound on $\Omega_{R}$.

This result then led to the idea that barriers with different values of $U_{b,max}$ could be applied to different members of an array of double-target BECs. The subsequent transfer behavior could place both lower and upper bounds on $\Omega_{R}$ constituting a measurement of this quantity and achieving the goal of the sensor.

We tested this idea by developing a ``design'' of a $2\times2$ double-target BEC array using our results for the $1\times1$ array.  The expected sensitivity range was defined and the expected outcomes of the sensor operation were described and associated with the bound values of $\Omega_{R}$. We simulated the operation of this designed $2\times2$ double-target BEC array. We found that this simulation agreed with expectations.  This supports proof-of-concept idea for the feasibility of our proposed rotation sensor.

In future work we intend to move beyond the proof-of-concept to explore which ranges of rotating-frame speed can be detected with this method and how sensitive a given double-target BEC is to changes in $\Omega_{R}$.  These studies will look at effects of changing the number of condensate atoms, the ring geometry, other characteristics of the target potential, and the shape and turn-on schedule of the barrier potential. We will also consider barriers with other characteristics that may be needed to access particular rotating-frame speed ranges.

One issue that might arise is how to tell the difference between rotation and acceleration when the sensor is far from the rotation axis. In Appendix\,\ref{rot_studies} we show that, as long as the distance from  the sensor to rotation axis is much larger than the spatial extent of the sensor, the sensor behavior is independent of this distance. However, when this condition is satisfied, it may be difficult to distinguish between linear and rotational acceleration of the sensor. 

One way to do this might be to design a sensor which has two identical double-target-array BECs instead of one as described above. In this scenario, in the initialization step one of the arrays is imprinted with clockwise circulation while the other is imprinted with counterclockwise circulation. For linear acceleration, the behavior of the two arrays would be the same and could be rejected in common mode. For rotational acceleration their behavior would be differential.  This question will be addressed in a future publication.

\begin{acknowledgments}
The authors wish to thank Cass Sackett, Gerhard Birkl, Nick Proukakis, Tom Bland, and Alex Yakimenko for stimulating and valuable discussions. This work was supported by the U.S. National Science Foundation under Grant No.\,PHY-2207476 and by the Physics Frontier Center under Grant No.\,PHY-1430094.  The authors also wish to acknowledge support from the National Institute of Standards and Technology.
\end{acknowledgments}

\appendix

\section{Simulation Model}
\label{sim_model}

The model that we used to simulate the behavior of a Bose-Einstein condensate is the 2D rotating-frame, time-dependent Gross-Pitaevskii equation (RFGPE). This equation is derived by starting with the 3D Gross-Pitaevskii equation which is the equation of motion for the condensate wave function in the non-rotating frame (which we will call the primed frame and the rotating frame will be the unprimed frame) and has the form
\begin{equation}
i\hbar
\frac{\partial\Psi^{\prime}}{\partial t^{\prime}} = 
-\frac{\hbar^{2}}{2M}\nabla^{\prime 2}\Psi^{\prime} +
V_{\rm ext}({\bf r}^{\prime},t^{\prime})\Psi^{\prime} + 
g_{3D}N\left|\Psi^{\prime}\right|^{2}\Psi^{\prime} .
\label{3dgpe}
\end{equation}
This is called the 3D Gross-Pitaevskii equation (3D GPE).  The factor $g_{3D} = 4\pi\hbar^{2}a_{s}/M$, where $a_{s}$ is the $s$-wave scattering length.  The scattering length measures the strength of the binary scattering of ultracold atoms and depends on which atomic species is present in the gas.

Next we assume that the external potential, $V_{\rm ext}$, is composed of a harmonic oscillator potential along the $z^{\prime}$ axis plus an arbitrary space- and time-dependent trap potential in the $x^{\prime}y^{\prime}$ plane:
\begin{equation}
V_{\rm ext}(x^{\prime},y^{\prime},z^{\prime},t^{\prime}) = 
\frac{1}{2}M\omega_{z}^{2}z^{\prime 2} +
V_{\rm trap}(x^{\prime},y^{\prime},t^{\prime}).
\end{equation}
We further assume that $\omega_{z}$ is large enough so that the system is strongly squeezed in the vertical direction, thus confining it to a quasi-2D horizontal plane.  

By assuming that the solution of the 3D GPE is approximately the product of the ground-state wave function of the $z^{\prime}$ harmonic oscillator and an arbitrary 2D wave function,
\begin{equation}
\Psi^{\prime}(x^{\prime},y^{\prime},z^{\prime},t^{\prime}) = 
\phi_{0}^{\prime}(z^{\prime})\psi^{\prime}(x^{\prime},y^{\prime},t^{\prime})e^{-i\omega_{z}t^{\prime}/2} 
\label{2dansatz}
\end{equation}
where $\phi_{0}^{\prime}(z^{\prime})$ satisfies
\begin{equation}
-\frac{\hbar^{2}}{2M}\frac{d^{2}\phi_{0}^{\prime}}{dz^{\prime 2}} +
\tfrac{1}{2}M\omega_{z}^{2}z^{\prime 2}\phi_{0}^{\prime} = \tfrac{1}{2}\hbar\omega_{z}\phi_{0}^{\prime},
\end{equation}
we can convert our 3D GPE (Eq.\,\ref{3dgpe}) into an effective 2D GPE. This yields the 2D GPE in the non-rotating frame
\begin{equation}
i\hbar\frac{\partial\psi^{\prime}}{\partial t^{\prime}} = 
-\frac{\hbar^{2}}{2M}\nabla_{\perp}^{\prime 2}\psi^{\prime} + 
V_{\rm trap}(x^{\prime},y^{\prime},t^{\prime})\psi^{\prime} + 
g_{2D}N\left|\psi^{\prime}\right|^{2}\psi^{\prime}.
\label{nrf2dgpe}
\end{equation}
The symbol $\nabla_{\perp}^{2}$ is the 2D Laplacian operator.  The factor $g_{2D}$ is the renormalized atom-atom scattering strength:
\begin{equation}
g_{2D} = 
g_{3D}
\left(
\int_{-\infty}^{+\infty}\,dz
\left|\phi_{0}(z)\right|^{4}
\right).
\end{equation}

Finally we transform Eq.\,(\ref{nrf2dgpe}) to the frame that is rotating with respect to this non-rotating frame at speed $\Omega_{R}$. We use unprimed symbols to denote quantities measured with respect to this rotating frame.  The result of this is the 2D rotating-frame Gross-Pitaevskii equation which we use in all of our simulations.  This equation reads
\begin{eqnarray}
i\hbar\frac{\partial\psi}{\partial t} 
&=& 
-\frac{\hbar^{2}}{2M}\nabla_{\perp}^{2}\psi + 
V_{\rm trap}(x,y,t)\psi + 
g_{2D}N\left|\psi\right|^{2}\psi\nonumber\\ 
&-&
\Omega_{R}\hat{L}_{z}\psi(x,y,t)
\label{2drfgpe}
\end{eqnarray}
where $\hat{L}_{z}$ is the $z$-component of the angular momentum operator:
\begin{equation}
\hat{L}_{z} = 
\frac{\hbar}{i}
\left(
x\frac{\partial}{\partial y} -
y\frac{\partial}{\partial x}
\right).
\end{equation}

The main quantities that will we used to analyze our simulations were the condensate velocity distribution given by
\begin{equation}
{\bf v}(x,y,t) = 
\frac{\hbar}{M}
{\bf \nabla}_{\perp}\theta(x,y,t),
\label{phase_grad}
\end{equation}
and the winding number, $n_{w}$, around a given closed path ${\cal C}$ given by
\begin{equation}
\Gamma_{\cal C} = 
\oint_{\cal C} 
{\bf v}\cdot\,d{\bf l} =
\left(\frac{2\pi\hbar}{M}\right)n_{w}.
\end{equation}
In all of our simulations we numerically computed the above integral for closed paths around the midtrack of the various ring BECs that are present in the double-target-BEC array.  If there was no flow in the ring, then $n_{w}=0$ and, if there was one unit of flow, then $n_{w}=1$.

\section{Target, double-target, and barrier potentials}
\label{potentials}

Here we give precise definitions of the potentials used in our simulations.  The full 2D trap potential can be written as the sum of the double-target-array potential plus the barrier potential. Thus, at point ${\bf r}\equiv (x,y)$,
\begin{equation}
V_{\rm trap}({\bf r},t) = 
V_{\rm dtap}({\bf r}) + 
V_{\rm barrier}({\bf r},t)
\end{equation}
The initial condensate is formed in the presence $V_{\rm dtap}({\bf r})$ while $V_{\rm barrier}({\bf r},t)$ is off. The barrier is turned on after the condensate formation and phase imprint steps.

A target potential has a disk whose center is located at coordinates ${\bf r}_{c}\equiv (x_{c},y_{c})$ and whose width is denoted by $w_{d}$.  The disk is surrounded by a concentric annulus whose midtrack radius is $R_{r}$ and whose width is $w_{r}$. The lowest value of the potential is at the center of the well and is very close to zero.  The highest value is reached outside the ring and has a height denoted by $V_{0}$.  The mathematical form of this potential is given by a sum of Gaussians~\cite{PhysRevX.4.031052}:
\begin{eqnarray}
V_{\rm tp}
(V_{0},{\bf r}_{c},w_{d},R_{r},w_{r};{\bf r}) 
&=&
V_{0}
\Bigg[
1 - 
\exp
\left(
-\left(\frac{\rho({\bf r})}{w_{d}}\right)^{2}
\right)\nonumber\\
&-&
\exp
\left(
-\left(
\frac
{\rho({\bf r})-R_{r}}
{w_{r}}
\right)^{2}
\right)
\Bigg]\nonumber\\
\end{eqnarray}
where
\begin{equation}
\rho({\bf r}) = 
\left(
(x-x_{c})^{2} + (y-y_{c})^{2}
\right)^{1/2}
\label{tp}
\end{equation}
With this definition we can now define the double-target potential.

The double-target potential is two target potentials with an overlap region.  We define the center of the double-target potential to be in the center of this overlap region and denote the coordinates of this point as ${\bf r}_{dt}\equiv (x_{dt},y_{dt})$.  The mathematical form of the double-target potential is given by
\begin{eqnarray}
V_{\rm dtp}({\bf r}_{dt};{\bf r})
&=&
V_{\rm tp}
(V_{0},x_{dt},y_{dt}+R_{r},w_{d},R_{r},w_{r};{\bf r})\nonumber\\
&\times&
\Theta(y_{dt}-\rho({\bf r}))
\nonumber\\
&+&
V_{\rm tp}
(V_{0},x_{dt},y_{dt}-R_{r},w_{d},R_{r},w_{r};{\bf r})\nonumber\\
&\times&
\Theta(\rho({\bf r})-y_{dt}),
\end{eqnarray}
where $\Theta(x)$ is the Heaviside step function. The first target-potential term above is the ``top ring'' whose center $y$ coordinate is higher than the double-target center by the ring radius, $R_{r}$. The second term is the ``bottom ring'' whose center $y$ coordinate is lower than the double-target center by $R_{r}$. The step
function is used to cut off the contribution of the bottom-ring target potential in the top-ring region and vice versa.  Far from the double-target center, the contribution of this potential is equal to $V_{0}$.

The double-target array potential is a lattice of $N_{rows}$ rows and $N_{cols}$ columns of double-target potentials. The $x$ coordinates of the double-target centers are located at coordinates $x_{i}$ where $i=1,\dots,N_{cols}$ and the $y$ coordinates are located at $y_{j}$ where $j=1,\dots,N_{rows}$.  The formula for this potential is
\begin{eqnarray}
V_{\rm dtap}(x,y) 
&=& 
\sum_{i}^{N_{cols}}
\sum_{j}^{N_{rows}}
V_{\rm dtp}(x_{i},y_{j};x,y)\\
&-&
V_{0}\left(N_{rows}N_{cols}-1\right).
\end{eqnarray}
The last factor subtracts off the contribution of all of the other double-target potentials at the site of a given double-target.

Finally we treat the barrier potential.  The barrier potential for a double-target array is just a sum of 2D Gaussian potentials, each one centered at one of the double-target centers.  Furthermore, the barrier at the double-target whose center is $(x_{i},y_{j})$ has a unique maximum barrier height, denoted by $U_{b,max,ij}$.  The barrier potential is given by 
\begin{eqnarray}
V_{\rm barrier}(x,y,t) 
&=& f(t)
\sum_{i}^{N_{cols}}
\sum_{j}^{N_{rows}}
U_{b,max,ij}\nonumber\\
&\times&
\exp
\left\{
-(x-x_{i})^{2}/w_{\parallel}^{2}
\right\}\nonumber\\
&\times&
\exp
\left\{
-(y-y_{i})^{2}/w_{\perp}^{2}
\right\}
\end{eqnarray}
where $w_{\parallel}$ and $w_{\perp}$ are the Gaussian widths along the $x$ and $y$ axes, respectively. The factor $f(t)$ is the time-dependent turn-on/hold/turn-off envelope that is shown in Fig.\,\ref{barrier_on_off}.  All of the individual barriers have the same time-dependent factor, $f(t)$.

\section{Rotation axis studies}
\label{rot_studies}

We conducted a set of simulations where a barrier was turned on and off across the overlap region of a $1\times1$ double-target BEC array.  In all simulations the conditions were as described in Section\,\ref{simulations}, and the value of $U_{b,max}$ was 45.7 nK.

\begin{figure}
\centering
\includegraphics[scale=0.15]{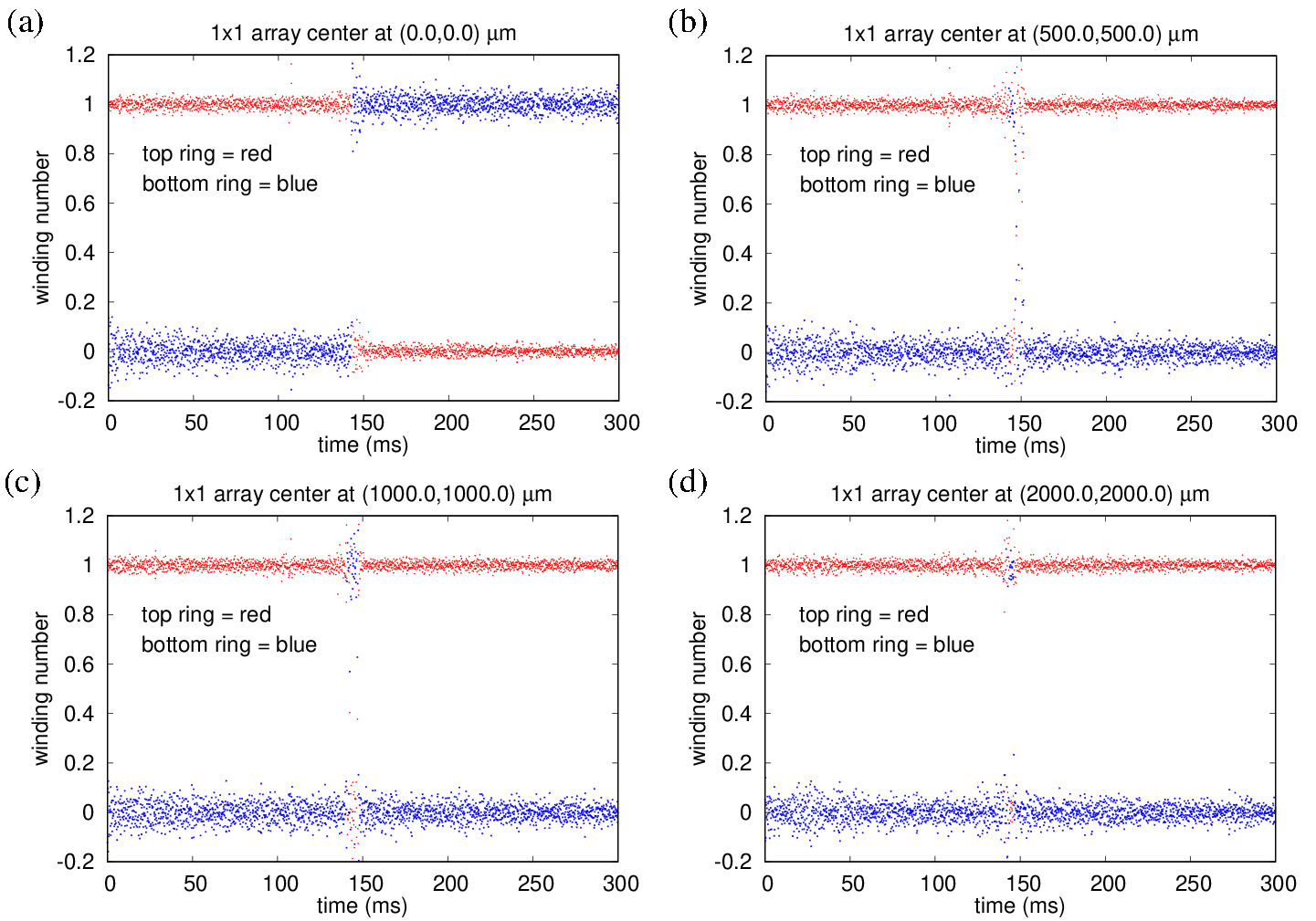}
\caption{Winding number versus time for the $1\times1$ double-target BEC array located at different distances from the rotation axis. (a) array center at the rotation axis $(x_{c},y_{c})=(0,0)\,\mu$m, (b) array center at $(x_{c},y_{c})=(500,500)\,\mu$m, (c) array center at $(x_{c},y_{c})=(1000,1000)\,\mu$m, (d) array center at $(x_{c},y_{c})=(2000,2000)\,\mu$m.}
\label{1x1_rot_results}
\end{figure}

The results of these simulations are illustrated in Fig.\,\ref{1x1_rot_results}. In each panel we show a plot of the winding number of the top ring (red) and the bottom ring (blue) as a function of time since the beginning of the barrier turn on.  The conditions of each of these simulations differs only in the location of the array center coordinates.  These are (a) array center at the rotation axis $(x_{c},y_{c})=(0,0)\,\mu$m, (b) array center at $(x_{c},y_{c})=(500,500)\,\mu$m, (c) array center at $(x_{c},y_{c})=(1000,1000)\,\mu$m, (d) array center at $(x_{c},y_{c})=(2000,2000)\,\mu$m. 

Figure\,\ref{1x1_rot_results}(a) shows that, when the array is centered on the rotation axis, that the flow transfers during the barrier turn on/off.  The winding numbers start at $t=0$ with the red curve on top and blue on bottom.  At the end ($t=300$ ms) this situation is reversed and so the flow transferred.  

On the other hand, when the array center is moved to $(x_{c},y_{c})=(500,500)\,\mu$m and shown in Fig.\,\ref{1x1_rot_results}(b), the red and blue curves do not reverse.  This indicates that there was no transfer. When the array center at $(x_{c},y_{c})=(1000,1000)\,\mu$m  and at $(x_{c},y_{c})=(2000,2000)\,\mu$m (Figs.\,\ref{1x1_rot_results}(c) and (d)) we see that the results are the same as for $(x_{c},y_{c})=(500,500)\,\mu$m.  In particular, the plots in Figs.\,\ref{1x1_rot_results}(c) and (d) are very similar to each other.

These results suggest that, when the array is close to the rotation axis, the transfer behavior can depend on the distance to the rotation axis.  \textcolor{black}{However, when the distance from rotation axis to array center is large compared to the size of the array, the transfer behavior stabilizes and the behavior becomes independent of the distance. Here the array size is roughly $100\times 100\,\mu$m.} We conducted a similar study with a $2\times 2$ array with identical results.

\bibliography{dtap}

\begin{thebibliography}{34}
\expandafter\ifx\csname natexlab\endcsname\relax\def\natexlab#1{#1}\fi
\expandafter\ifx\csname bibnamefont\endcsname\relax
  \def\bibnamefont#1{#1}\fi
\expandafter\ifx\csname bibfnamefont\endcsname\relax
  \def\bibfnamefont#1{#1}\fi
\expandafter\ifx\csname citenamefont\endcsname\relax
  \def\citenamefont#1{#1}\fi
\expandafter\ifx\csname url\endcsname\relax
  \def\url#1{\texttt{#1}}\fi
\expandafter\ifx\csname urlprefix\endcsname\relax\def\urlprefix{URL }\fi
\providecommand{\bibinfo}[2]{#2}
\providecommand{\eprint}[2][]{\url{#2}}

\bibitem[{\citenamefont{Grewal et~al.}(2013)\citenamefont{Grewal, Andrews, and
  Bartone}}]{navbook}
\bibinfo{author}{\bibfnamefont{M.}~\bibnamefont{Grewal}},
  \bibinfo{author}{\bibfnamefont{A.}~\bibnamefont{Andrews}}, \bibnamefont{and}
  \bibinfo{author}{\bibfnamefont{C.}~\bibnamefont{Bartone}},
  \emph{\bibinfo{title}{Global Navigation, Satellite Systems, Intertial
  Navigation, and Integration}} (\bibinfo{publisher}{John Wiley and Sons},
  \bibinfo{address}{Hoboken, New Jersey}, \bibinfo{year}{2013}).

\bibitem[{\citenamefont{Pai and Marakala}(2016)}]{ins_conference}
\bibinfo{author}{\bibfnamefont{K.~R.} \bibnamefont{Pai}} \bibnamefont{and}
  \bibinfo{author}{\bibfnamefont{N.}~\bibnamefont{Marakala}}, in
  \emph{\bibinfo{booktitle}{2016 International Conference on Electrical,
  Electronics, and Optimization Techniques}} (\bibinfo{publisher}{IEEE},
  \bibinfo{year}{2016}), pp. \bibinfo{pages}{1682--1686}, ISBN
  \bibinfo{isbn}{978-1-4673-9939-5}.

\bibitem[{\citenamefont{El-Sheimy and Youssef}(2020)}]{El-Sheimy}
\bibinfo{author}{\bibfnamefont{N.}~\bibnamefont{El-Sheimy}} \bibnamefont{and}
  \bibinfo{author}{\bibfnamefont{A.}~\bibnamefont{Youssef}},
  \bibinfo{journal}{Satellite Navigation} \textbf{\bibinfo{volume}{1}},
  \bibinfo{pages}{2} (\bibinfo{year}{2020}), ISSN \bibinfo{issn}{2662-1363},
  \urlprefix\url{https://doi.org/10.1186/s43020-019-0001-5}.

\bibitem[{\citenamefont{Pasienski and DeMarco}(2008)}]{demarco_2008}
\bibinfo{author}{\bibfnamefont{M.}~\bibnamefont{Pasienski}} \bibnamefont{and}
  \bibinfo{author}{\bibfnamefont{B.}~\bibnamefont{DeMarco}},
  \bibinfo{journal}{Optical Express} \textbf{\bibinfo{volume}{16}},
  \bibinfo{pages}{2176} (\bibinfo{year}{2008}).

\bibitem[{\citenamefont{Gaunt and Hadzibabic}(2008)}]{hadzibabic_2012}
\bibinfo{author}{\bibfnamefont{A.}~\bibnamefont{Gaunt}} \bibnamefont{and}
  \bibinfo{author}{\bibfnamefont{Z.}~\bibnamefont{Hadzibabic}},
  \bibinfo{journal}{Scientific Reports} \textbf{\bibinfo{volume}{2}},
  \bibinfo{pages}{721} (\bibinfo{year}{2008}).

\bibitem[{\citenamefont{Henderson et~al.}(2009)\citenamefont{Henderson, Ryu,
  MacCormick, and Boshier}}]{boshier_2009}
\bibinfo{author}{\bibfnamefont{K.}~\bibnamefont{Henderson}},
  \bibinfo{author}{\bibfnamefont{C.}~\bibnamefont{Ryu}},
  \bibinfo{author}{\bibfnamefont{C.}~\bibnamefont{MacCormick}},
  \bibnamefont{and} \bibinfo{author}{\bibfnamefont{M.}~\bibnamefont{Boshier}},
  \bibinfo{journal}{New Journal of Physics} \textbf{\bibinfo{volume}{11}},
  \bibinfo{pages}{043030} (\bibinfo{year}{2009}).

\bibitem[{\citenamefont{Bowman et~al.}(2017)\citenamefont{Bowman, Harte,
  Chardonnet, Groot, Denny, Goc, Anderson, Ireland, Cassettari, and
  Bruce}}]{donatella_cassettari}
\bibinfo{author}{\bibfnamefont{D.}~\bibnamefont{Bowman}},
  \bibinfo{author}{\bibfnamefont{T.~L.} \bibnamefont{Harte}},
  \bibinfo{author}{\bibfnamefont{V.}~\bibnamefont{Chardonnet}},
  \bibinfo{author}{\bibfnamefont{C.~D.} \bibnamefont{Groot}},
  \bibinfo{author}{\bibfnamefont{S.~J.} \bibnamefont{Denny}},
  \bibinfo{author}{\bibfnamefont{G.~L.} \bibnamefont{Goc}},
  \bibinfo{author}{\bibfnamefont{M.}~\bibnamefont{Anderson}},
  \bibinfo{author}{\bibfnamefont{P.}~\bibnamefont{Ireland}},
  \bibinfo{author}{\bibfnamefont{D.}~\bibnamefont{Cassettari}},
  \bibnamefont{and} \bibinfo{author}{\bibfnamefont{G.~D.} \bibnamefont{Bruce}},
  \bibinfo{journal}{Opt. Express} \textbf{\bibinfo{volume}{25}},
  \bibinfo{pages}{11692} (\bibinfo{year}{2017}),
  \urlprefix\url{http://www.opticsexpress.org/abstract.cfm?URI=oe-25-10-11692}.

\bibitem[{\citenamefont{Amico et~al.}(2017)\citenamefont{Amico, Birkl, Boshier,
  and Kwek}}]{Amico_2017}
\bibinfo{author}{\bibfnamefont{L.}~\bibnamefont{Amico}},
  \bibinfo{author}{\bibfnamefont{G.}~\bibnamefont{Birkl}},
  \bibinfo{author}{\bibfnamefont{M.}~\bibnamefont{Boshier}}, \bibnamefont{and}
  \bibinfo{author}{\bibfnamefont{L.-C.} \bibnamefont{Kwek}},
  \bibinfo{journal}{New Journal of Physics} \textbf{\bibinfo{volume}{19}},
  \bibinfo{pages}{020201} (\bibinfo{year}{2017}),
  \urlprefix\url{https://doi.org/10.1088%2F1367-2630%2Faa5a6d}.

\bibitem[{\citenamefont{Moan et~al.}(2020)\citenamefont{Moan, Horne,
  Arpornthip, Luo, Fallon, Berl, and Sackett}}]{PhysRevLett.124.120403}
\bibinfo{author}{\bibfnamefont{E.~R.} \bibnamefont{Moan}},
  \bibinfo{author}{\bibfnamefont{R.~A.} \bibnamefont{Horne}},
  \bibinfo{author}{\bibfnamefont{T.}~\bibnamefont{Arpornthip}},
  \bibinfo{author}{\bibfnamefont{Z.}~\bibnamefont{Luo}},
  \bibinfo{author}{\bibfnamefont{A.~J.} \bibnamefont{Fallon}},
  \bibinfo{author}{\bibfnamefont{S.~J.} \bibnamefont{Berl}}, \bibnamefont{and}
  \bibinfo{author}{\bibfnamefont{C.~A.} \bibnamefont{Sackett}},
  \bibinfo{journal}{Phys. Rev. Lett.} \textbf{\bibinfo{volume}{124}},
  \bibinfo{pages}{120403} (\bibinfo{year}{2020}),
  \urlprefix\url{https://link.aps.org/doi/10.1103/PhysRevLett.124.120403}.

\bibitem[{\citenamefont{{Gustavson} et~al.}(2000)\citenamefont{{Gustavson},
  {Landragin}, and {Kasevich}}}]{2000Gustavson}
\bibinfo{author}{\bibfnamefont{T.~L.} \bibnamefont{{Gustavson}}},
  \bibinfo{author}{\bibfnamefont{A.}~\bibnamefont{{Landragin}}},
  \bibnamefont{and} \bibinfo{author}{\bibfnamefont{M.~A.}
  \bibnamefont{{Kasevich}}}, \bibinfo{journal}{Classical and Quantum Gravity}
  \textbf{\bibinfo{volume}{17}}, \bibinfo{pages}{2385} (\bibinfo{year}{2000}).

\bibitem[{\citenamefont{Ryu et~al.}(2013)\citenamefont{Ryu, Blackburn, Blinova,
  and Boshier}}]{boshier_2013a}
\bibinfo{author}{\bibfnamefont{C.}~\bibnamefont{Ryu}},
  \bibinfo{author}{\bibfnamefont{P.~W.} \bibnamefont{Blackburn}},
  \bibinfo{author}{\bibfnamefont{A.~A.} \bibnamefont{Blinova}},
  \bibnamefont{and} \bibinfo{author}{\bibfnamefont{M.~G.}
  \bibnamefont{Boshier}}, \bibinfo{journal}{Phys. Rev. Lett.}
  \textbf{\bibinfo{volume}{111}}, \bibinfo{pages}{205301}
  (\bibinfo{year}{2013}),
  \urlprefix\url{https://link.aps.org/doi/10.1103/PhysRevLett.111.205301}.

\bibitem[{\citenamefont{Mathey and Mathey}(2016)}]{Mathey2016}
\bibinfo{author}{\bibfnamefont{A.~C.} \bibnamefont{Mathey}} \bibnamefont{and}
  \bibinfo{author}{\bibfnamefont{L.}~\bibnamefont{Mathey}},
  \bibinfo{journal}{New Journal of Physics} \textbf{\bibinfo{volume}{18}},
  \bibinfo{pages}{055016} (\bibinfo{year}{2016}),
  \urlprefix\url{http://stacks.iop.org/1367-2630/18/i=5/a=055016}.

\bibitem[{\citenamefont{Wang et~al.}(2015)\citenamefont{Wang, Kumar,
  Jendrzejewski, Wilson, Edwards, Eckel, Campbell, and Clark}}]{njp_paper}
\bibinfo{author}{\bibfnamefont{Y.-H.} \bibnamefont{Wang}},
  \bibinfo{author}{\bibfnamefont{A.}~\bibnamefont{Kumar}},
  \bibinfo{author}{\bibfnamefont{F.}~\bibnamefont{Jendrzejewski}},
  \bibinfo{author}{\bibfnamefont{R.~M.} \bibnamefont{Wilson}},
  \bibinfo{author}{\bibfnamefont{M.}~\bibnamefont{Edwards}},
  \bibinfo{author}{\bibfnamefont{S.}~\bibnamefont{Eckel}},
  \bibinfo{author}{\bibfnamefont{G.~K.} \bibnamefont{Campbell}},
  \bibnamefont{and} \bibinfo{author}{\bibfnamefont{C.~W.} \bibnamefont{Clark}},
  \bibinfo{journal}{New Journal of Physics} \textbf{\bibinfo{volume}{17}},
  \bibinfo{pages}{125012} (\bibinfo{year}{2015}),
  \urlprefix\url{http://stacks.iop.org/1367-2630/17/i=12/a=125012}.

\bibitem[{\citenamefont{Safavi-Naini et~al.}(2016)\citenamefont{Safavi-Naini,
  Capogrosso-Sansone, Kuklov, and Penna}}]{qs7}
\bibinfo{author}{\bibfnamefont{A.}~\bibnamefont{Safavi-Naini}},
  \bibinfo{author}{\bibfnamefont{B.}~\bibnamefont{Capogrosso-Sansone}},
  \bibinfo{author}{\bibfnamefont{A.}~\bibnamefont{Kuklov}}, \bibnamefont{and}
  \bibinfo{author}{\bibfnamefont{V.}~\bibnamefont{Penna}},
  \bibinfo{journal}{New Journal of Physics} \textbf{\bibinfo{volume}{18}},
  \bibinfo{pages}{025017} (\bibinfo{year}{2016}),
  \urlprefix\url{http://stacks.iop.org/1367-2630/18/i=2/a=025017}.

\bibitem[{\citenamefont{Bell et~al.}(2016)\citenamefont{Bell, Glidden, Humbert,
  Bromley, Haine, Davis, Neely, Baker, and Rubinsztein-Dunlop}}]{Bell2016}
\bibinfo{author}{\bibfnamefont{T.~A.} \bibnamefont{Bell}},
  \bibinfo{author}{\bibfnamefont{J.~A.~P.} \bibnamefont{Glidden}},
  \bibinfo{author}{\bibfnamefont{L.}~\bibnamefont{Humbert}},
  \bibinfo{author}{\bibfnamefont{M.~W.~J.} \bibnamefont{Bromley}},
  \bibinfo{author}{\bibfnamefont{S.~A.} \bibnamefont{Haine}},
  \bibinfo{author}{\bibfnamefont{M.~J.} \bibnamefont{Davis}},
  \bibinfo{author}{\bibfnamefont{T.~W.} \bibnamefont{Neely}},
  \bibinfo{author}{\bibfnamefont{M.~A.} \bibnamefont{Baker}}, \bibnamefont{and}
  \bibinfo{author}{\bibfnamefont{H.}~\bibnamefont{Rubinsztein-Dunlop}},
  \bibinfo{journal}{New Journal of Physics} \textbf{\bibinfo{volume}{18}},
  \bibinfo{pages}{035003} (\bibinfo{year}{2016}),
  \urlprefix\url{http://stacks.iop.org/1367-2630/18/i=3/a=035003}.

\bibitem[{\citenamefont{Murray et~al.}(2013)\citenamefont{Murray, Krygier,
  Edwards, Wright, Campbell, and Clark}}]{ring_circ_probe}
\bibinfo{author}{\bibfnamefont{N.}~\bibnamefont{Murray}},
  \bibinfo{author}{\bibfnamefont{M.}~\bibnamefont{Krygier}},
  \bibinfo{author}{\bibfnamefont{M.}~\bibnamefont{Edwards}},
  \bibinfo{author}{\bibfnamefont{K.~C.} \bibnamefont{Wright}},
  \bibinfo{author}{\bibfnamefont{G.~K.} \bibnamefont{Campbell}},
  \bibnamefont{and} \bibinfo{author}{\bibfnamefont{C.~W.} \bibnamefont{Clark}},
  \bibinfo{journal}{Phys. Rev. A} \textbf{\bibinfo{volume}{88}},
  \bibinfo{pages}{053615} (\bibinfo{year}{2013}),
  \urlprefix\url{https://link.aps.org/doi/10.1103/PhysRevA.88.053615}.

\bibitem[{\citenamefont{Gallucci and Proukakis}(2016)}]{sq3}
\bibinfo{author}{\bibfnamefont{D.}~\bibnamefont{Gallucci}} \bibnamefont{and}
  \bibinfo{author}{\bibfnamefont{N.~P.} \bibnamefont{Proukakis}},
  \bibinfo{journal}{New Journal of Physics} \textbf{\bibinfo{volume}{18}},
  \bibinfo{pages}{025004} (\bibinfo{year}{2016}),
  \urlprefix\url{http://stacks.iop.org/1367-2630/18/i=2/a=025004}.

\bibitem[{\citenamefont{Aghamalyan et~al.}(2016)\citenamefont{Aghamalyan,
  Nguyen, Auksztol, Gan, Valado, Condylis, Kwek, Dumke, and Amico}}]{sq7}
\bibinfo{author}{\bibfnamefont{D.}~\bibnamefont{Aghamalyan}},
  \bibinfo{author}{\bibfnamefont{N.~T.} \bibnamefont{Nguyen}},
  \bibinfo{author}{\bibfnamefont{F.}~\bibnamefont{Auksztol}},
  \bibinfo{author}{\bibfnamefont{K.~S.} \bibnamefont{Gan}},
  \bibinfo{author}{\bibfnamefont{M.~M.} \bibnamefont{Valado}},
  \bibinfo{author}{\bibfnamefont{P.~C.} \bibnamefont{Condylis}},
  \bibinfo{author}{\bibfnamefont{L.-C.} \bibnamefont{Kwek}},
  \bibinfo{author}{\bibfnamefont{R.}~\bibnamefont{Dumke}}, \bibnamefont{and}
  \bibinfo{author}{\bibfnamefont{L.}~\bibnamefont{Amico}},
  \bibinfo{journal}{New Journal of Physics} \textbf{\bibinfo{volume}{18}},
  \bibinfo{pages}{075013} (\bibinfo{year}{2016}),
  \urlprefix\url{http://stacks.iop.org/1367-2630/18/i=7/a=075013}.

\bibitem[{\citenamefont{Navez et~al.}(2016)\citenamefont{Navez, Pandey, Mas,
  Poulios, Fernholz, and von Klitzing}}]{sq8}
\bibinfo{author}{\bibfnamefont{P.}~\bibnamefont{Navez}},
  \bibinfo{author}{\bibfnamefont{S.}~\bibnamefont{Pandey}},
  \bibinfo{author}{\bibfnamefont{H.}~\bibnamefont{Mas}},
  \bibinfo{author}{\bibfnamefont{K.}~\bibnamefont{Poulios}},
  \bibinfo{author}{\bibfnamefont{T.}~\bibnamefont{Fernholz}}, \bibnamefont{and}
  \bibinfo{author}{\bibfnamefont{W.}~\bibnamefont{von Klitzing}},
  \bibinfo{journal}{New Journal of Physics} \textbf{\bibinfo{volume}{18}},
  \bibinfo{pages}{075014} (\bibinfo{year}{2016}),
  \urlprefix\url{http://stacks.iop.org/1367-2630/18/i=7/a=075014}.

\bibitem[{\citenamefont{Ramanathan et~al.}(2011)\citenamefont{Ramanathan,
  Wright, Muniz, Zelan, Hill, Lobb, Helmerson, Phillips, and
  Campbell}}]{1st_ringBEC_current}
\bibinfo{author}{\bibfnamefont{A.}~\bibnamefont{Ramanathan}},
  \bibinfo{author}{\bibfnamefont{K.~C.} \bibnamefont{Wright}},
  \bibinfo{author}{\bibfnamefont{S.~R.} \bibnamefont{Muniz}},
  \bibinfo{author}{\bibfnamefont{M.}~\bibnamefont{Zelan}},
  \bibinfo{author}{\bibfnamefont{W.~T.} \bibnamefont{Hill}},
  \bibinfo{author}{\bibfnamefont{C.~J.} \bibnamefont{Lobb}},
  \bibinfo{author}{\bibfnamefont{K.}~\bibnamefont{Helmerson}},
  \bibinfo{author}{\bibfnamefont{W.~D.} \bibnamefont{Phillips}},
  \bibnamefont{and} \bibinfo{author}{\bibfnamefont{G.~K.}
  \bibnamefont{Campbell}}, \bibinfo{journal}{Phys. Rev. Lett.}
  \textbf{\bibinfo{volume}{106}}, \bibinfo{pages}{130401}
  (\bibinfo{year}{2011}),
  \urlprefix\url{https://link.aps.org/doi/10.1103/PhysRevLett.106.130401}.

\bibitem[{\citenamefont{Wright et~al.}(2013)\citenamefont{Wright, Blakestad,
  Lobb, Phillips, and Campbell}}]{2nd_ringBEC_current}
\bibinfo{author}{\bibfnamefont{K.~C.} \bibnamefont{Wright}},
  \bibinfo{author}{\bibfnamefont{R.~B.} \bibnamefont{Blakestad}},
  \bibinfo{author}{\bibfnamefont{C.~J.} \bibnamefont{Lobb}},
  \bibinfo{author}{\bibfnamefont{W.~D.} \bibnamefont{Phillips}},
  \bibnamefont{and} \bibinfo{author}{\bibfnamefont{G.~K.}
  \bibnamefont{Campbell}}, \bibinfo{journal}{Phys. Rev. Lett.}
  \textbf{\bibinfo{volume}{110}}, \bibinfo{pages}{025302}
  (\bibinfo{year}{2013}),
  \urlprefix\url{https://link.aps.org/doi/10.1103/PhysRevLett.110.025302}.

\bibitem[{\citenamefont{Kumar et~al.}(2016)\citenamefont{Kumar, Anderson,
  Phillips, Eckel, Campbell, and Stringari}}]{Kumar2016}
\bibinfo{author}{\bibfnamefont{A.}~\bibnamefont{Kumar}},
  \bibinfo{author}{\bibfnamefont{N.}~\bibnamefont{Anderson}},
  \bibinfo{author}{\bibfnamefont{W.~D.} \bibnamefont{Phillips}},
  \bibinfo{author}{\bibfnamefont{S.}~\bibnamefont{Eckel}},
  \bibinfo{author}{\bibfnamefont{G.~K.} \bibnamefont{Campbell}},
  \bibnamefont{and}
  \bibinfo{author}{\bibfnamefont{S.}~\bibnamefont{Stringari}},
  \bibinfo{journal}{New Journal of Physics} \textbf{\bibinfo{volume}{18}},
  \bibinfo{pages}{025001} (\bibinfo{year}{2016}),
  \urlprefix\url{http://stacks.iop.org/1367-2630/18/i=2/a=025001}.

\bibitem[{\citenamefont{Eckel et~al.}(2014{\natexlab{a}})\citenamefont{Eckel,
  Lee, Jendrzejewski, Murray, Clark, Lobb, Phillips, Edwards, and
  Campbell}}]{hysteresis_nature_paper}
\bibinfo{author}{\bibfnamefont{S.}~\bibnamefont{Eckel}},
  \bibinfo{author}{\bibfnamefont{J.~G.} \bibnamefont{Lee}},
  \bibinfo{author}{\bibfnamefont{F.}~\bibnamefont{Jendrzejewski}},
  \bibinfo{author}{\bibfnamefont{N.}~\bibnamefont{Murray}},
  \bibinfo{author}{\bibfnamefont{C.~W.} \bibnamefont{Clark}},
  \bibinfo{author}{\bibfnamefont{C.~J.} \bibnamefont{Lobb}},
  \bibinfo{author}{\bibfnamefont{W.~D.} \bibnamefont{Phillips}},
  \bibinfo{author}{\bibfnamefont{M.}~\bibnamefont{Edwards}}, \bibnamefont{and}
  \bibinfo{author}{\bibfnamefont{G.~K.} \bibnamefont{Campbell}},
  \bibinfo{journal}{Nature} \textbf{\bibinfo{volume}{506}},
  \bibinfo{pages}{200} (\bibinfo{year}{2014}{\natexlab{a}}).

\bibitem[{\citenamefont{Jendrzejewski et~al.}(2014)\citenamefont{Jendrzejewski,
  Eckel, Murray, Lanier, Edwards, Lobb, and Campbell}}]{resistive_flow_BEC}
\bibinfo{author}{\bibfnamefont{F.}~\bibnamefont{Jendrzejewski}},
  \bibinfo{author}{\bibfnamefont{S.}~\bibnamefont{Eckel}},
  \bibinfo{author}{\bibfnamefont{N.}~\bibnamefont{Murray}},
  \bibinfo{author}{\bibfnamefont{C.}~\bibnamefont{Lanier}},
  \bibinfo{author}{\bibfnamefont{M.}~\bibnamefont{Edwards}},
  \bibinfo{author}{\bibfnamefont{C.~J.} \bibnamefont{Lobb}}, \bibnamefont{and}
  \bibinfo{author}{\bibfnamefont{G.~K.} \bibnamefont{Campbell}},
  \bibinfo{journal}{Phys. Rev. Lett.} \textbf{\bibinfo{volume}{113}},
  \bibinfo{pages}{045305} (\bibinfo{year}{2014}),
  \urlprefix\url{https://link.aps.org/doi/10.1103/PhysRevLett.113.045305}.

\bibitem[{\citenamefont{Bland et~al.}(2020)\citenamefont{Bland, Marolleau,
  Comaron, Malomed, and Proukakis}}]{Bland_2020}
\bibinfo{author}{\bibfnamefont{T.}~\bibnamefont{Bland}},
  \bibinfo{author}{\bibfnamefont{Q.}~\bibnamefont{Marolleau}},
  \bibinfo{author}{\bibfnamefont{P.}~\bibnamefont{Comaron}},
  \bibinfo{author}{\bibfnamefont{B.~A.} \bibnamefont{Malomed}},
  \bibnamefont{and} \bibinfo{author}{\bibfnamefont{N.~P.}
  \bibnamefont{Proukakis}}, \bibinfo{journal}{Journal of Physics B: Atomic,
  Molecular and Optical Physics} \textbf{\bibinfo{volume}{53}},
  \bibinfo{pages}{115301} (\bibinfo{year}{2020}),
  \urlprefix\url{https://dx.doi.org/10.1088/1361-6455/ab81e9}.

\bibitem[{\citenamefont{Bland et~al.}(2022)\citenamefont{Bland, Yatsuta,
  Edwards, Nikolaieva, Oliinyk, Yakimenko, and
  Proukakis}}]{PhysRevResearch.4.043171}
\bibinfo{author}{\bibfnamefont{T.}~\bibnamefont{Bland}},
  \bibinfo{author}{\bibfnamefont{I.~V.} \bibnamefont{Yatsuta}},
  \bibinfo{author}{\bibfnamefont{M.}~\bibnamefont{Edwards}},
  \bibinfo{author}{\bibfnamefont{Y.~O.} \bibnamefont{Nikolaieva}},
  \bibinfo{author}{\bibfnamefont{A.~O.} \bibnamefont{Oliinyk}},
  \bibinfo{author}{\bibfnamefont{A.~I.} \bibnamefont{Yakimenko}},
  \bibnamefont{and} \bibinfo{author}{\bibfnamefont{N.~P.}
  \bibnamefont{Proukakis}}, \bibinfo{journal}{Phys. Rev. Res.}
  \textbf{\bibinfo{volume}{4}}, \bibinfo{pages}{043171} (\bibinfo{year}{2022}),
  \urlprefix\url{https://link.aps.org/doi/10.1103/PhysRevResearch.4.043171}.

\bibitem[{\citenamefont{Chaika et~al.}(2024)\citenamefont{Chaika, Oliinyk,
  Yatsuta, Proukakis, Edwards, Yakimenko, and Bland}}]{chaika2024}
\bibinfo{author}{\bibfnamefont{A.}~\bibnamefont{Chaika}},
  \bibinfo{author}{\bibfnamefont{A.~O.} \bibnamefont{Oliinyk}},
  \bibinfo{author}{\bibfnamefont{I.~V.} \bibnamefont{Yatsuta}},
  \bibinfo{author}{\bibfnamefont{N.~P.} \bibnamefont{Proukakis}},
  \bibinfo{author}{\bibfnamefont{M.}~\bibnamefont{Edwards}},
  \bibinfo{author}{\bibfnamefont{A.~I.} \bibnamefont{Yakimenko}},
  \bibnamefont{and} \bibinfo{author}{\bibfnamefont{T.}~\bibnamefont{Bland}},
  \emph{\bibinfo{title}{Acceleration-induced transport of quantum vortices in
  joined atomtronic circuits}} (\bibinfo{year}{2024}), \eprint{2410.23818},
  \urlprefix\url{https://arxiv.org/abs/2410.23818}.

\bibitem[{\citenamefont{Pethick and Smith}(2008)}]{pethick_smith_2008}
\bibinfo{author}{\bibfnamefont{C.~J.} \bibnamefont{Pethick}} \bibnamefont{and}
  \bibinfo{author}{\bibfnamefont{H.}~\bibnamefont{Smith}},
  \emph{\bibinfo{title}{Bose–Einstein Condensation in Dilute Gases}}
  (\bibinfo{publisher}{Cambridge University Press}, \bibinfo{year}{2008}),
  \bibinfo{edition}{2nd} ed.

\bibitem[{\citenamefont{Gauthier et~al.}(2016)\citenamefont{Gauthier, Lenton,
  Parry, Baker, Davis, Rubinsztein-Dunlop, and Neely}}]{dmd_reference}
\bibinfo{author}{\bibfnamefont{G.}~\bibnamefont{Gauthier}},
  \bibinfo{author}{\bibfnamefont{I.}~\bibnamefont{Lenton}},
  \bibinfo{author}{\bibfnamefont{N.~M.} \bibnamefont{Parry}},
  \bibinfo{author}{\bibfnamefont{M.}~\bibnamefont{Baker}},
  \bibinfo{author}{\bibfnamefont{M.~J.} \bibnamefont{Davis}},
  \bibinfo{author}{\bibfnamefont{H.}~\bibnamefont{Rubinsztein-Dunlop}},
  \bibnamefont{and} \bibinfo{author}{\bibfnamefont{T.~W.} \bibnamefont{Neely}},
  \bibinfo{journal}{Optica} \textbf{\bibinfo{volume}{3}}, \bibinfo{pages}{1136}
  (\bibinfo{year}{2016}),
  \urlprefix\url{http://opg.optica.org/optica/abstract.cfm?URI=optica-3-10-1136}.

\bibitem[{\citenamefont{Amico et~al.}(2021)\citenamefont{Amico, Boshier, Birkl,
  Minguzzi, Miniatura, Kwek, Aghamalyan, Ahufinger, Anderson, Andrei
  et~al.}}]{roadmap}
\bibinfo{author}{\bibfnamefont{L.}~\bibnamefont{Amico}},
  \bibinfo{author}{\bibfnamefont{M.}~\bibnamefont{Boshier}},
  \bibinfo{author}{\bibfnamefont{G.}~\bibnamefont{Birkl}},
  \bibinfo{author}{\bibfnamefont{A.}~\bibnamefont{Minguzzi}},
  \bibinfo{author}{\bibfnamefont{C.}~\bibnamefont{Miniatura}},
  \bibinfo{author}{\bibfnamefont{L.-C.} \bibnamefont{Kwek}},
  \bibinfo{author}{\bibfnamefont{D.}~\bibnamefont{Aghamalyan}},
  \bibinfo{author}{\bibfnamefont{V.}~\bibnamefont{Ahufinger}},
  \bibinfo{author}{\bibfnamefont{D.}~\bibnamefont{Anderson}},
  \bibinfo{author}{\bibfnamefont{N.}~\bibnamefont{Andrei}},
  \bibnamefont{et~al.}, \bibinfo{journal}{AVS Quantum Science}
  \textbf{\bibinfo{volume}{3}}, \bibinfo{pages}{039201} (\bibinfo{year}{2021}),
  \eprint{https://doi.org/10.1116/5.0026178},
  \urlprefix\url{https://doi.org/10.1116/5.0026178}.

\bibitem[{\citenamefont{Eckel et~al.}(2014{\natexlab{b}})\citenamefont{Eckel,
  Jendrzejewski, Kumar, Lobb, and Campbell}}]{PhysRevX.4.031052}
\bibinfo{author}{\bibfnamefont{S.}~\bibnamefont{Eckel}},
  \bibinfo{author}{\bibfnamefont{F.}~\bibnamefont{Jendrzejewski}},
  \bibinfo{author}{\bibfnamefont{A.}~\bibnamefont{Kumar}},
  \bibinfo{author}{\bibfnamefont{C.~J.} \bibnamefont{Lobb}}, \bibnamefont{and}
  \bibinfo{author}{\bibfnamefont{G.~K.} \bibnamefont{Campbell}},
  \bibinfo{journal}{Phys. Rev. X} \textbf{\bibinfo{volume}{4}},
  \bibinfo{pages}{031052} (\bibinfo{year}{2014}{\natexlab{b}}),
  \urlprefix\url{https://link.aps.org/doi/10.1103/PhysRevX.4.031052}.

\bibitem[{\citenamefont{Eckel et~al.}(2018)\citenamefont{Eckel, Kumar,
  Jacobson, Spielman, and Campbell}}]{exp_universe}
\bibinfo{author}{\bibfnamefont{S.}~\bibnamefont{Eckel}},
  \bibinfo{author}{\bibfnamefont{A.}~\bibnamefont{Kumar}},
  \bibinfo{author}{\bibfnamefont{T.}~\bibnamefont{Jacobson}},
  \bibinfo{author}{\bibfnamefont{I.~B.} \bibnamefont{Spielman}},
  \bibnamefont{and} \bibinfo{author}{\bibfnamefont{G.~K.}
  \bibnamefont{Campbell}}, \bibinfo{journal}{Phys. Rev. X}
  \textbf{\bibinfo{volume}{8}}, \bibinfo{pages}{021021} (\bibinfo{year}{2018}),
  \urlprefix\url{https://link.aps.org/doi/10.1103/PhysRevX.8.021021}.

\bibitem[{\citenamefont{Gauthier et~al.}(2021)\citenamefont{Gauthier, Bell,
  Stilgoe, Baker, Rubinsztein-Dunlop, and Neely}}]{GAUTHIER20211}
\bibinfo{author}{\bibfnamefont{G.}~\bibnamefont{Gauthier}},
  \bibinfo{author}{\bibfnamefont{T.~A.} \bibnamefont{Bell}},
  \bibinfo{author}{\bibfnamefont{A.~B.} \bibnamefont{Stilgoe}},
  \bibinfo{author}{\bibfnamefont{M.}~\bibnamefont{Baker}},
  \bibinfo{author}{\bibfnamefont{H.}~\bibnamefont{Rubinsztein-Dunlop}},
  \bibnamefont{and} \bibinfo{author}{\bibfnamefont{T.~W.} \bibnamefont{Neely}},
  in \emph{\bibinfo{booktitle}{Dynamic high-resolution optical trapping of
  ultracold atoms}}, edited by \bibinfo{editor}{\bibfnamefont{L.~F.}
  \bibnamefont{Dimauro}},
  \bibinfo{editor}{\bibfnamefont{H.}~\bibnamefont{Perrin}}, \bibnamefont{and}
  \bibinfo{editor}{\bibfnamefont{S.~F.} \bibnamefont{Yelin}}
  (\bibinfo{publisher}{Academic Press}, \bibinfo{year}{2021}),
  vol.~\bibinfo{volume}{70} of \emph{\bibinfo{series}{Advances In Atomic,
  Molecular, and Optical Physics}}, pp. \bibinfo{pages}{1--101},
  \urlprefix\url{https://www.sciencedirect.com/science/article/pii/S1049250X2100001X}.

\bibitem[{\citenamefont{Mathew et~al.}(2015)\citenamefont{Mathew, Kumar, Eckel,
  Jendrzejewski, Campbell, Edwards, and Tiesinga}}]{spirals_paper}
\bibinfo{author}{\bibfnamefont{R.}~\bibnamefont{Mathew}},
  \bibinfo{author}{\bibfnamefont{A.}~\bibnamefont{Kumar}},
  \bibinfo{author}{\bibfnamefont{S.}~\bibnamefont{Eckel}},
  \bibinfo{author}{\bibfnamefont{F.}~\bibnamefont{Jendrzejewski}},
  \bibinfo{author}{\bibfnamefont{G.~K.} \bibnamefont{Campbell}},
  \bibinfo{author}{\bibfnamefont{M.}~\bibnamefont{Edwards}}, \bibnamefont{and}
  \bibinfo{author}{\bibfnamefont{E.}~\bibnamefont{Tiesinga}},
  \bibinfo{journal}{Phys. Rev. A} \textbf{\bibinfo{volume}{92}},
  \bibinfo{pages}{033602} (\bibinfo{year}{2015}),
  \urlprefix\url{https://link.aps.org/doi/10.1103/PhysRevA.92.033602}.

\end{thebibliography}

\end{document}